\documentclass[twocolumn,showpacs]{revtex4}

\usepackage[latin1]{inputenc}
\usepackage{graphicx}%
\usepackage{dcolumn}
\usepackage{amsmath}

\makeatletter
\def\btt#1{\texttt{\@backslashchar#1}}%
\DeclareRobustCommand\bblash{\btt{\@backslashchar}}%
\makeatother

\begin{document}

\title{Microscopic Theory of Traffic Flow Instability Governing Traffic Breakdown at Highway Bottlenecks:
Growing Wave of Increase in Speed  in Synchronized Flow}

\mark{Microscopic Theory of Instability Governing Traffic Breakdown at Highway Bottlenecks}

\author{Boris S. Kerner$^1$}

 \affiliation{$^1$
Physics of Transport and Traffic, University Duisburg-Essen,
47048 Duisburg, Germany}



\begin{abstract} 
We  have revealed a  
 growing  local  speed wave of   increase in    speed  that
can randomly occur in synchronized flow  (S) at a highway bottleneck. The development of such a traffic flow
  instability leads to free flow (F) at the bottleneck; therefore, we call this  instability as an S$\rightarrow$F instability.
 Whereas the S$\rightarrow$F instability leads to a local {\it increase in   speed} (growing acceleration wave), in contrast,
the classical traffic flow instability introduced in 50s--60s and incorporated later in a huge number of traffic flow models
 leads to a growing wave of a local {\it  decrease in speed} (growing deceleration wave).
We have found that the S$\rightarrow$F instability can occur only, if there is a finite time delay in  driver over-acceleration.
The initial speed disturbance of increase in  speed  (called $\lq\lq$speed peak") that initiates the S$\rightarrow$F instability
occurs usually at the downstream front of synchronized flow   at the bottleneck.
There can be many speed peaks with random amplitudes that occur randomly over time. 
It has been found that the S$\rightarrow$F instability exhibits the nucleation nature: Only when a speed peak  amplitude  
is large enough, the S$\rightarrow$F instability occurs; in contrast, speed peaks of smaller amplitudes cause dissolving  speed  waves of a local increase in speed
 (dissolving  acceleration    waves)
in synchronized flow.  
We have found that the S$\rightarrow$F instability governs traffic breakdown --
a phase transition from   free flow to synchronized flow (F$\rightarrow$S transition)   at the bottleneck:
The nucleation nature of  the S$\rightarrow$F instability explains the metastability of free flow with respect to an F$\rightarrow$S transition
at the bottleneck. 
\end{abstract}

\maketitle

 \section {Introduction  \label{Int}} 

In 1958--1961, Herman, Gazis,  Montroll, Potts,   Rothery, and Chandler~\cite{GH195910,Gazis1961A10,GH10,Chandler} from General Motors (GM) Company revealed the existence of
a traffic flow instability associated with {\it a driver over-deceleration effect}:
If a vehicle begins
to decelerate unexpectedly, then due to a finite driver reaction time   the following vehicle  
 starts deceleration with a delay.    As a result,
  the   speed of the following vehicle  becomes lower than the speed of the preceding vehicle. If this over-deceleration effect
is realized for all following drivers, the traffic flow instability occurs leading to
  a growing wave of a local  {\it speed   decrease}  
    in traffic flow that can be considered $\lq\lq$growing deceleration wave" in traffic flow.  
With the use of very different mathematical approaches, this classical traffic flow instability 
has been incorporated in a huge number of traffic flow models; examples are well-known Kometani-Sasaki model~\cite{KS,KS1959A},
optimal velocity (OV) model by 
Newell~\cite{Newell1961,Newell1963A,Newell1981}, a stochastic version of Newell's model~\cite{Newell_Stoch},  Gipps model~\cite{Gipps,Gipps1986}, Wiedemann's model~\cite{Wiedemann}, 
Whitham's model~\cite{Whitham1990}, Payne's macroscopic model~\cite{ach_Pay197110,ach_Pay197910}, 
the Nagel-Schreckenberg (NaSch) cellular automaton (CA) model~\cite{Stoc},  the OV model
by Bando {\it et al.}~\cite{Bando1995}, a stochastic model by Krau{\ss} {\it et al.}~\cite{ach_Kra10}, a lattice model by Nagatani~\cite{fail_Nagatani1998A,fail_Nagatani1999B}, 
Treiber's intelligent driver model~\cite{ach_Helbing200010}, the Aw-Rascle macroscopic model~\cite{ach_Aw200010}, 
a full velocity difference OV model    by Jiang {\it et al.}~\cite{ach_Jiang2001A}
and a huge number of other traffic flow models 
(see references in     
 books and reviews~\cite{Reviews,Reviews2,Kerner_Review}). All these 
different traffic flow models
 can be considered belonging to the same GM model class. Indeed,
 as found firstly in 1993--1994~\cite{KK1993}, in all these very different traffic flow models 
the classical instability 
 leads to a moving jam (J) formation in free flow (F) (F$\rightarrow$J transition) 
 (see references in~\cite{Reviews2,KernerBook,KernerBook2,Kerner_Review}). 
 The classical instability  of the GM model class   should explain traffic breakdown, i.e.,
a transition from free flow to congested traffic observed in real 
traffic~\cite{GH195910,Gazis1961A10,GH10,Chandler,KS,KS1959A,Newell1961,Newell1963A,Newell1981,Newell_Stoch,Gipps,Gipps1986,Wiedemann,Whitham1990,ach_Pay197110,ach_Pay197910,Stoc,Bando1995,ach_Kra10,fail_Nagatani1998A,fail_Nagatani1999B,ach_Helbing200010,ach_Aw200010,ach_Jiang2001A,Reviews,Reviews2}).

\begin{figure}
\begin{center}
\includegraphics*[width=8 cm]{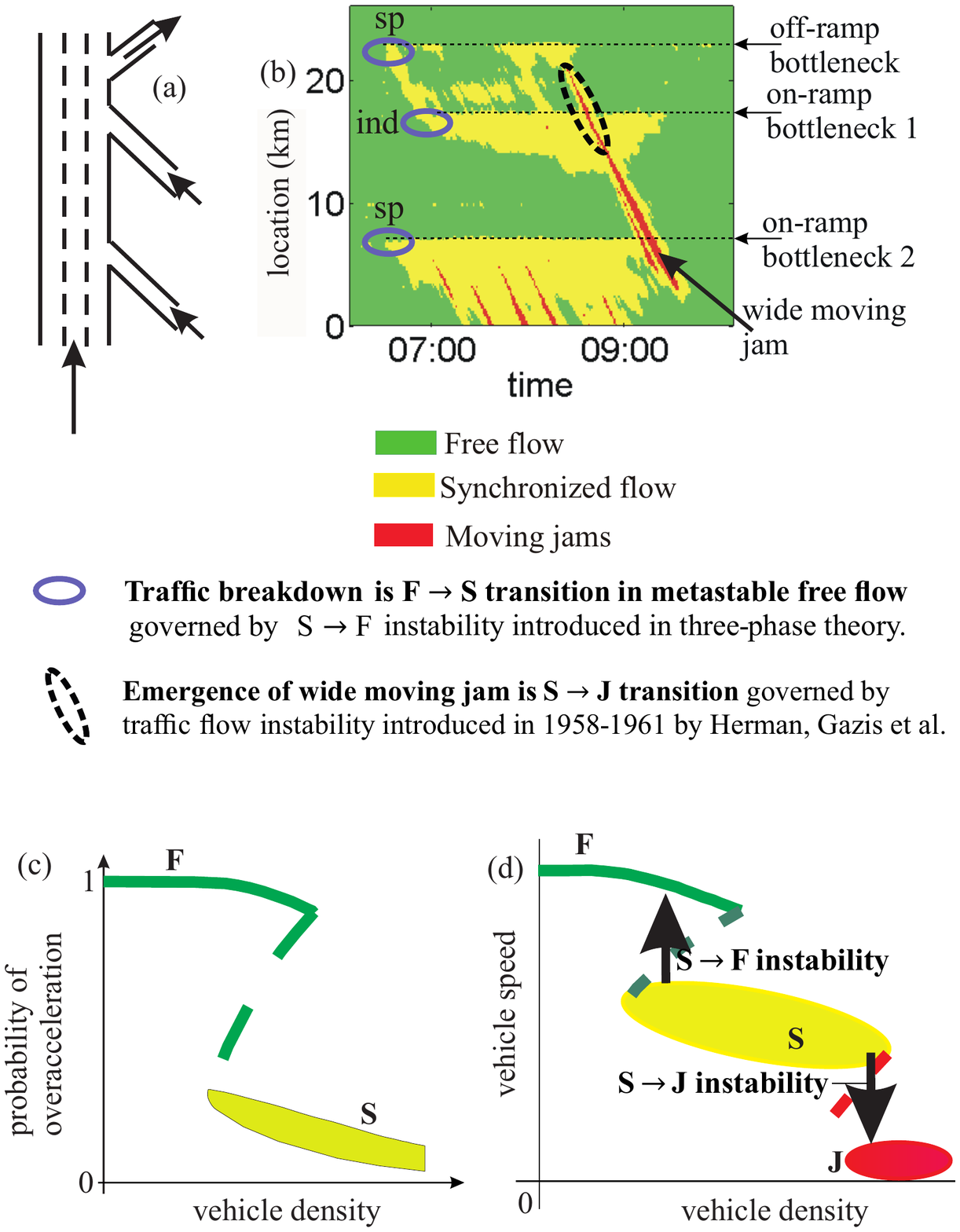}
\end{center}
\caption[]{A known empirical example of phase transitions in traffic flow illustrating two traffic flow instabilities 
of three-phase   theory (real measured traffic data of road detectors installed along three-lane highway) (a, b)~\cite{Kerner_2014} and illustrations of associated hypotheses of three-phase
 theory (c, d): (a) Sketch of 
   section of three-lane highway in Germany with three bottlenecks. (b)
Speed  data  measured with road detectors installed 
 along    road section in (a);   data~\cite{KernerBook} are
presented in space and time with   averaging   method described in Sec.~C.2 of~\cite{KernerRSch2013}.
(c) Hypothesis of three-phase theory about discontinuous character of over-acceleration probability~\cite{Kerner1999A,Kerner1999B,Kerner1999C,KernerBook,KernerBook2}.
(d) Hypothesis of three-phase theory about F$\rightarrow$S$\rightarrow$J  phase transitions in traffic flow: 2Z-characteristic for phase transitions~\cite{KernerBook,Kerner1999B}.
  F -- free flow phase, S -- synchronized flow phase, J -- wide moving jam phase. In (b), $\lq\lq$sp" -- spontaneous F$\rightarrow$S transition,
  $\lq\lq$ind" -- induced F$\rightarrow$S transition~\cite{KernerBook}. 
  }
\label{TwoIn_Emp} 
\end{figure}
 
 However, as shown in~\cite{KernerBook,KernerBook2,Kerner_Review}, traffic flow models
 models of the GM model class (see references in~\cite{Kerner_Review,KernerBook,KernerBook2})
failed in the explanation of real traffic breakdown. This is because rather than an F$\rightarrow$J transition
of the models of the GM model class, in all  real field traffic data traffic breakdown is
   a phase transition from a metastable free flow to synchronized flow
(F$\rightarrow$S transition)~\cite{KR1997,Kerner1997A7,Kerner1998B,Kerner1999A,Kerner1999B,Kerner1999C,KernerBook,KernerBook2,Kerner_Review,Kerner2002A,Waves,Kerner_Review2}.

To explain an F$\rightarrow$S transition  in metastable free flow,   a  three-phase traffic theory ($\lq\lq$three-phase theory" for short)
has been introduced~\cite{Kerner1997A7,Kerner1998B,Kerner1999A,Kerner1999B,Kerner1999C,KernerBook,KernerBook2,Kerner_Review} which in addition 
 to the free flow phase (F), there are two phases in congested traffic: 
the synchronized flow (S) and wide moving jam (J) phases.    One of  the characteristic features of the three-phase theory is the
  assumption about the existence of {\it two}
  qualitatively different instabilities in vehicular traffic: 
  
  (i) A traffic flow instability predicted in three-phase  theory~\cite{Kerner1999A,Kerner1999B,Kerner1999C,KernerBook,KernerBook2,Kerner_Review} that is
associated with  {\it an over-acceleration effect}. It is assumed that probability of over-acceleration  
should  
exhibit a {\it discontinuous character}~\cite{Kerner1999A,Kerner1999B,Kerner1999C,KernerBook,KernerBook2} (Fig.~\ref{TwoIn_Emp} (c)).  
 Due to the discontinuous character of the over-acceleration probability the  instability
  (labeled by S$\rightarrow$F instability in Fig.~\ref{TwoIn_Emp} (d)) should cause
  a growing wave of  a local  {\it increase} in the vehicle speed in synchronized flow. Respectively,
  in the three-phase theory it is assumed that a spatiotemporal competition between the over-acceleration effect
  and the speed adaptation effect occurring  in car-following
  leads to the metastability of free flow with respect to an F$\rightarrow$S transition  at the bottleneck. 
The assumption that traffic breakdown at a highway bottleneck is the F$\rightarrow$S transition occurring  
 in metastable free flow is the basic assumption of the three-phase theory~\cite{Kerner1999A,Kerner1999B,Kerner1999C,KernerBook,KernerBook2,Kerner_Review}.
  
  (ii) In the three-phase   theory it is further assumed that rather than traffic breakdown,
  the instability of the GM model class explains a phase transition from synchronized flow to wide moving jams
  (S$\rightarrow$J transition) that is labeled by S$\rightarrow$J instability in Fig.~\ref{TwoIn_Emp} (d).
  
 The first mathematical implementation of these hypotheses of three-phase theory~\cite{Kerner1997A7,Kerner1998B,Kerner1999A,Kerner1999B,Kerner1999C,KernerBook,KernerBook2} has been a stochastic
 continuous in space microscopic model~\cite{KKl} and a   CA  three-phase model~\cite{KKW}, which 
 has been further developed for 
 different applications in~\cite{KKl2003A,Kerner2008A,Kerner2008B,Kerner2008C,Kerner2008D,KKl2009A,Kerner_2014,KKl2004A,Kerner_Hyp,KKHS2013,KKS2014A,Heavy,KKS2011,KKl2006AA,Kerner_EPL,Kerner_Diff}.
 Over time there has been developed 
a number of other 
three-phase     flow
models   (e.g.,~\cite{Davis,Lee_Sch2004A,Jiang2004A,Gao2007,Davis2006,Davis2006b,Davis2006d,Davis2006e,Davis2010,Davis2011,Jiang2007A,Jiang2005A,Jiang2005B,Jiang2007C,Pott2007A,Li,Wu2008,Laval2007A8,Hoogendoorn20088,Wu2009,Jia2009,Tian2009,He2009,Jin2010,Klenov,Klenov2,Kokubo,LeeKim2011,Jin2011,Neto2011,Zhang2011,Wei-Hsun2011IEEEA,Lee2011A,Tian2012,Kimathi2012B,Wang2012A,Tian2012B,Qiu2013,YangLu2013A,KnorrSch2013A,XiangZhengTao2013A,Mendez2013A,Rui2014A,Hausken2015A,Tian2015A,Rui2015A,Rui2015B,Rui2015C,Rui2015D,Xu2015A,Davis2015A}) that incorporate some of the  hypotheses of the three-phase 
theory~\cite{Kerner1999A,Kerner1999B,Kerner1999C,KernerBook,KernerBook2}.

  The hypothesis that the S$\rightarrow$F instability  at a highway bottleneck   should govern the nucleation nature of an F$\rightarrow$S transition, i.e., the metastability of free flow
with respect to an F$\rightarrow$S transition (traffic breakdown)  was introduced in the three-phase theory  many 
years ago~\cite{Kerner1999A,Kerner1999B,Kerner1999C,KernerBook} (Fig.~\ref{TwoIn_Emp} (d)). However,   microscopic physical features of this S$\rightarrow$F instability
have been unknown up to now. In particular, the following theoretical questions arise, which have {\it not} been
answered in earlier theoretical studies of three-phase  flow 
models~\cite{KernerBook,KernerBook2,KKl,KKl2003A,Kerner2008A,Kerner2008B,Kerner2008C,Kerner2008D,KKl2009A,Kerner_2014,KKl2004A,KKHS2013,KKS2014A,Heavy,KKS2011,KKW,KKl2006AA,Kerner_Diff}:

(i) What is a  disturbance in synchronized flow   that can spontaneously initiate the S$\rightarrow$F instability  at the bottleneck?

(ii) Can be proven that the S$\rightarrow$F instability  at the bottleneck exhibits the nucleation nature? 

(iii) How does the S$\rightarrow$F instability occurring in {\it synchronized flow} governs 
the  metastability of {\it free flow}  with respect to the F$\rightarrow$S transition at the bottleneck? Indeed, in accordance with
in the three-phase theory~\cite{KernerBook}  the speed adaptation effect, which   describes the tendency from free flow to synchronized
flow, cannot lead to some traffic flow instability.
Therefore, the speed adaptation effect cannot be the origin of the nucleation nature of  the F$\rightarrow$S transition at the bottleneck
observed in real traffic.

(iv) What is the physics of a random time delay to the F$\rightarrow$S transition at the bottleneck found in
simulations with stochastic three-phase traffic flow models~\cite{KernerBook,KernerBook2,KKl2003A,Kerner2008A,Kerner2008B,Kerner2008C,Kerner2008D,KKl2009A,Kerner_2014,KKl2004A,KKHS2013,KKS2014A,Heavy,KKS2011,KKW}?

   In this article, we reveal  microscopic features of    the S$\rightarrow$F instability that answer the above questions (i)--(iv).
   We will show that this microscopic theory of the S$\rightarrow$F instability exhibits a general character: All results can be derived with
   very different mathematical stochastic three-phase traffic flow models, in particular with the KKSW (Kerner-Klenov-Schreckenberg-Wolf)
    CA model~\cite{KKW,KKHS2013,KKS2014A} and the Kerner-Klenov stochastic 
   model~\cite{KKl,KKl2003A,KKl2009A,Kerner_2014,KKl2004A,Heavy}. Because the KKSW CA model is considerably more simple one than the Kerner-Klenov stochastic 
   model, we present results of the microscopic theory of the S$\rightarrow$F instability based on a study on the KKSW CA model; associated results derived with 
   the Kerner-Klenov stochastic 
   model are briefly considered in discussion section.
  
  The article is organized as follows. In Sec.~\ref{Inst_Mic_Over_I_S}, we show the existence of an
  S$\rightarrow$F   instability at a highway bottleneck. The nucleation nature of an S$\rightarrow$F   instability    at the bottleneck is the subject
  of Sec.~\ref{Dis_Wave_KKSW_S}.
   Microscopic features of a random time-delayed   traffic breakdown (F$\rightarrow$S  transition) at highway bottlenecks are studied in Sec.~\ref{Nuc_Sec}. This analysis
   proves that the S$\rightarrow$F   instability governs traffic breakdown at the bottleneck. A general character of this conclusion
   is shown   in Sec.~\ref{Gen_S}.
  In   Sec.~\ref{Dis_S}, we compare  the classical traffic flow instability of the GM model class with the S$\rightarrow$F   instability
   of three-phase theory (Sec.~\ref{GM_S}),
 discuss   cases in which either there is no over-acceleration in the KKSW CA model (Sec.~\ref{No_Over_KKSW_S})
or  there is no time delay in over-acceleration in the KKSW CA model (Sec.~\ref{NoDelay_Over_KKSW_S}), make a generalization of the results based on an analysis of the Kerner-Klenov stochastic 
   model (Sec.~\ref{KKSW_KKl_S})
as well as formulate conclusions (Sec.~\ref{Con_S}).

  \section{S$\rightarrow$F Traffic Flow Instability   \label{Inst_Mic_Over_I_S}}
 
   \subsection{KKSW CA Model}
   
   To study the S$\rightarrow$F traffic flow instability in   synchronized flow   at a highway bottleneck,  we use 
the KKSW CA three-phase traffic flow model~\cite{KKW,KKHS2013,KKS2014A} whose parameters are the same as those in~\cite{KKS2014A}.

        \begin{figure}
\begin{center}
\includegraphics*[width=8 cm]{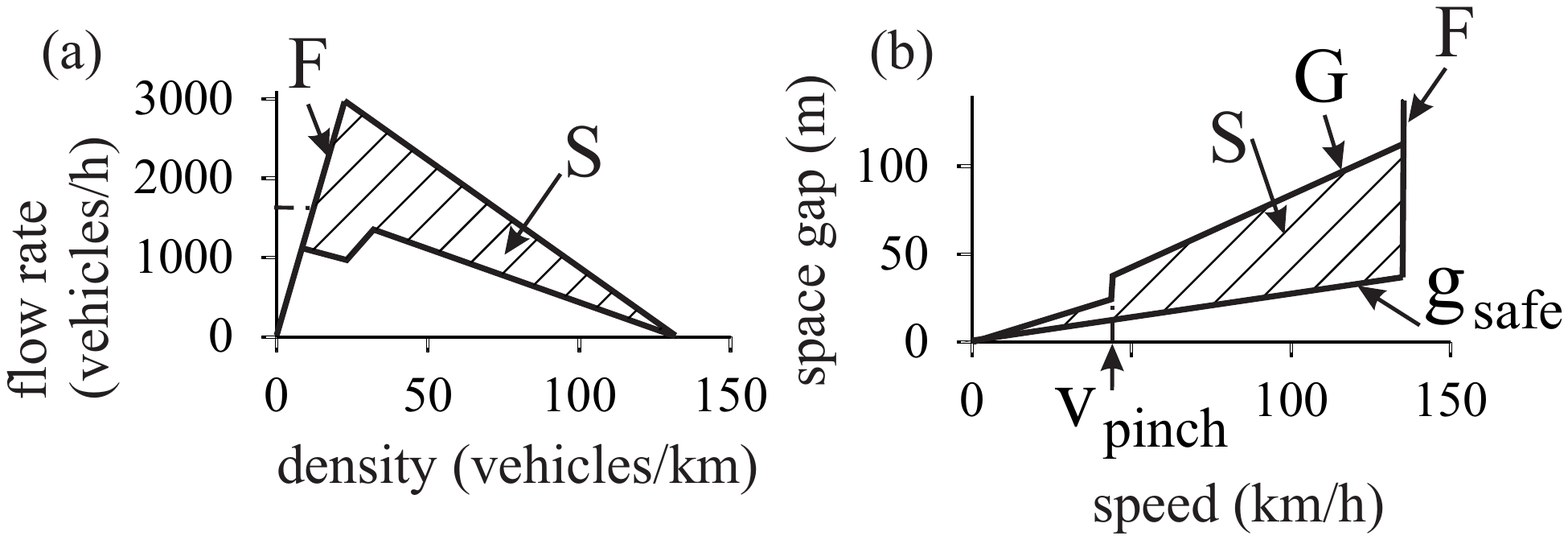}
\end{center}
\caption[]{Steady states of the KKSW CA model       in the flow--density
  (a) and space-gap--speed planes (b).  $G$ and $g_{\rm safe}$ are, respectively, a synchronization gap and
   a safe gap at a time-independent speed $v$ (where $g_{\rm safe}=v$), F -- free flow, S --
synchronized flow (hatched 2D-regions in (a, b)). Parameters of the KKSW CA model  
 used in simulations are as follows:  $d=$ 5 (7.5 m), 
$v_{\rm free}=$ 25 (135 km/h),  
 $p_3=$ 0.01, $p^{(2)}_{0}=$ 0.5,   
$v_{\rm pinch}=$ 8 (43.2 km/h), 
$k_1=$ 3, $k_2=$ 2.   
$p_{\rm a,1}=$0.07, $p_{\rm a,2}=$0.08, 
$p^{(2)}_{2}=$0.35,   
 $v_{\rm syn}=$14  (75.6 km/h), $\Delta v_{\rm syn}=$3  (16.2 km/h).
  }
\label{SteadyStates}
\end{figure}

\subsubsection{Rules of vehicle motion in KKSW CA model}
 
 In
 the KKSW CA model  for identical drivers and vehicles moving on   a single-lane road~\cite{KKS2014A}, the following  designations for main variables and vehicle parameters are used:
 $n=0, 1, 2, ...$ is the number of time steps; $\tau=1$ s is time step; $\delta x=1.5$ m is space step;
$x_{n}$ and $v_{n}$ are the coordinate and speed of the vehicle; time and space are measured in units of $\tau$ and $\delta x$, respectively;
$v_{\rm free}$ is the maximum speed in free flow; 
$g_{n}=x_{\ell,  n}-x_{n}-d$ is a space gap between two  vehicles following each other;
the lower index $\ell$ marks variables related to the preceding vehicle; $d$ is vehicle length; 
 $G_{n}$  is a synchronization space gap (Fig.~\ref{SteadyStates} (a, b)).

 The KKSW CA model consists of the following sequence of rules~\cite{KKS2014A}:
\begin{description} 
\item {(a)} $\lq\lq$comparison of vehicle gap with the synchronization gap":  
 \begin{eqnarray} 
\label{conadaptation1}
\quad \mbox{if} \ g_{n} \leq G(v_{n}) \nonumber \\ 
 \quad \mbox{then  follow rules (b), (c) and skip rule (d),} 
\end{eqnarray}
 \begin{eqnarray}
\label{nonadaptation1}
 \quad \mbox{if} \ g_{n} > G(v_{n}) \nonumber \\
 \quad \mbox{then skip rules (b), (c) and follow rule (d),}
\end{eqnarray}
\item {(b)} $\lq\lq$speed adaptation within   synchronization   gap" is given by formula:
\begin{equation}
\label{adaptation_KKSW}
v_{n+1}=v_{n}+\mathrm{sgn}(v_{\ell,  n}-v_{n}),
\end{equation}
\item {(c)} $\lq\lq$over-acceleration through random acceleration within   synchronization   gap"
is given by formula
  \begin{eqnarray}
\label{Overacceleration1_KKW}
\mbox{if}   
  \ v_{n} \geq v_{\ell,  n}, \ \mbox{then
with probability} \ p_{\rm a}, \nonumber \\   
\quad  v_{n+1}=\min(v_{n+1}+1, \ v_{\rm free}),
\end{eqnarray}
\item {(d)} $\lq\lq$acceleration":
   \begin{equation}
\label{acceleration_KKSW}
v_{n+1}=\min(v_{n}+1, \ v_{\rm free}),
\end{equation}
\item {(e)} $\lq\lq$deceleration":
\begin{equation}
\label{Deceleration_KKSW}
v_{n+1}=\min (v_{n+1}, \ g_{n}),  
\end{equation}
\item {(f)} $\lq\lq$randomization"  is given by formula:
\begin{equation}
\label{Randomization_KKSW}
{\rm with} \ {\rm probability} \ p, \quad v_{n+1}=\max (v_{n+1}-1, \ 0),
\end{equation}
\item {(g)} $\lq\lq$motion" is described by formula:
\begin{equation}
\label{Motion_KKSW}
x_{n+1}=x_{n}+ v_{n+1}.
\end{equation}
\end{description}
Formula (\ref{Overacceleration1_KKW})  
 is applied,
when 
\begin{equation}
  r<p_{\rm a},
\label{rand_p1}
\end{equation}
formula (\ref{Randomization_KKSW})  
 is applied,
when 
\begin{equation}
p_{\rm a} \leq r<p_{\rm a}+p, 
\label{rand_p}
\end{equation}
where $p_{\rm a}+p\leq 1$; $r=rand()$ is a random value distributed uniformly between 0 and 1.
Probability of over-acceleration $p_{\rm a}$ in (\ref{Overacceleration1_KKW}) is chosen as 
the increasing speed function:
\begin{equation}
p_{\rm a}(v_{n})=p_{\rm a,1}+p_{\rm a,2}\max (0, \min (1, \ (v_{n}-v_{\rm syn})/\Delta v_{\rm syn})),
\label{p_acc}
\end{equation}
where $p_{\rm a,1}$, $p_{\rm a,2}$, $v_{\rm syn}$ and $\Delta v_{\rm syn}$ are constants.
 In  (\ref{conadaptation1}), (\ref{nonadaptation1}),
\begin{equation}
G(v_{n})=kv_{n}.
\label{S_Gap}
\end{equation}

The rules of vehicle motion  (\ref{nonadaptation1})--(\ref{S_Gap}) (without formula   (\ref{p_acc}))
have been formulated in the KKW (Kerner-Klenov-Wolf) CA model~\cite{KKW}.
In comparison with the KKW CA model~\cite{KKW}, we use in
 (\ref{Randomization_KKSW}), (\ref{rand_p})
  for probability $p$ formula  
 \begin{equation}
p=\left\{
\begin{array}{ll}
p_{2} &  \textrm{for  $v_{n+1}>v_{n}$}, \\
p_{3} &  \textrm{for  $v_{n+1} \leq v_{n}$},
\end{array} \right.
\label{p_MSP}
\end{equation}
which has been used in the KKSW CA model of Ref.~\cite{KKHS2013}.
The importance of formula (\ref{p_MSP}) is as follows.
This  rule of vehicle motion leads to a time delay in vehicle acceleration at the downstream front of synchronized flow.
In other words, this is an additional mechanism of time delay in vehicle acceleration in comparison with 
 a well-known  slow-to-start rule~\cite{Stoc2,Schadschneider_Book}:
\begin{equation}
p_{2}(v_{n})=\left\{
\begin{array}{ll}
p^{(2)}_{0} &  \textrm{for $v_{n}=0$},\\
p^{(2)}_{1} &  \textrm{for $v_{n}>0$} \\ 
\end{array} \right. 
\label{general}
\end{equation}
that is also used in the KKSW CA model. However, in the KKSW CA model
in formula  (\ref{general})  
probability  $p^{(2)}_{1}$ is   chosen to provide a delay in vehicle acceleration
only if the vehicle does not accelerate at previous time step $n$:
\begin{equation}
p^{(2)}_{1}=\left\{
\begin{array}{ll}
p^{(2)}_{2} &  \textrm{for $v_{n} \leq v_{n-1}$}, \\ 
0 &  \textrm{for $v_{n} > v_{n-1}$}. \\ 
\end{array} \right. 
\label{p_delay}
\end{equation} 
In (\ref{p_MSP})--(\ref{p_delay}), $p_{3}$, $p^{(2)}_{0}$, and $p^{(2)}_{2}$ are constants.
We also assume that
in (\ref{S_Gap})~\cite{KKW}
\begin{equation}
k(v_{n})=\left\{
\begin{array}{ll}
k_{1} &  \textrm{for $v_{n} > v_{\rm pinch}$},\\
k_{2} &  \textrm{for $v_{n} \leq v_{\rm pinch}$}, \\ 
\end{array} \right. 
\label{Pinch}
\end{equation}
where $v_{\rm pinch}$, $k_{1}$, and $k_{2}$ are constants ($k_{1}>k_{2}\geq 1$).

The rule of vehicle motion (\ref{p_MSP}) of the KKSW CA model~\cite{KKHS2013} together with formula (\ref{p_acc}) allows us to improve 
characteristics of synchronized flow patterns (SP) simulated with the KKSW CA model (\ref{nonadaptation1})--(\ref{Pinch}) for a single-lane road. Other physical 
features of the KKSW CA model have been explained in~\cite{KKHS2013}.
A model of an on-ramp bottleneck is the same as that presented in~\cite{KKS2011}.

In accordance with qualitative three-phase theory~\cite{KernerBook}, a competition between  speed adaptation and over-acceleration 
should determine the existence of  an S$\rightarrow$F instability.
Thus it is useful to discuss the description of these effects with the KKSW CA model.

\subsubsection{Speed adaptation effect in KKSW CA model}

In the KKSW CA model, the speed adaptation effect in synchronized flow 
  takes place  within the space  gap range:
\begin{equation} 
g_{\rm safe, \ n}\leq g_{\rm n}\leq G_{\rm n}, 
\label{gap_range}
\end{equation}
where  $g_{\rm safe, \ n}$  is a safe space gap, $g_{\rm safe, \ n}=v_{n}$. Under  condition (\ref{gap_range}), formula (\ref{adaptation_KKSW}) is valid, i.e.,
the vehicle
  tends to adjust its speed to the preceding vehicle without caring, what the precise space gap is, as long as it is safe:
The vehicle accelerates or decelerates in dependence of whether the vehicle moves
slower or faster than the preceding vehicle, respectively. In other words, there are both $\lq\lq$negative" and $\lq\lq$positive"
 speed adaptation.

   \subsubsection{Time delay in over-acceleration in KKSW CA model} 
  
A formulation for model fluctuations that simulates over-acceleration on a single-lane road is as follows.
Each vehicle, which moves in synchronized flow with a space gap that satisfies   conditions (\ref{gap_range}) (Fig.~\ref{SteadyStates} (b)),
accelerates randomly with some probability $p_{\rm a}$ (\ref{Overacceleration1_KKW}). This random vehicle acceleration
occurs only under conditions 
  (\ref{gap_range}) and
\begin{equation}
v_{n} \geq v_{\ell,  n}.
\label{Over_formula}
\end{equation} 
Thus the vehicle accelerates with probability $p_{\rm a}$,
 even if the preceding vehicle does not accelerate and the vehicle speed is not lower
than the speed of the preceding vehicle.
Therefore, in accordance with the definition of   over-acceleration~\cite{KernerBook,KernerBook2}, this vehicle
acceleration is an example of over-acceleration. 
Because the probability of over-acceleration $p_{\rm a}<1$,
there is {\it on average} a time delay in   over-acceleration. The mean time delay in the over-acceleration is longer than time step 
of the KKSW CA model ($\tau=$ 1 s). 
The over-acceleration effect  results in the discontinuous character of the
probability of over-acceleration as a density (and flow rate) function
as   required by the associated hypothesis of three-phase theory~\cite{Kerner1999A,Kerner1999B,Kerner1999C,KernerBook,KernerBook2} (Fig.~\ref{TwoIn_Emp} (c)).

The probability of over-acceleration $p_{\rm a}$ (\ref{Overacceleration1_KKW}) is an increasing function of vehicle speed.
 This model feature
 supports the over-acceleration  within a local speed disturbance of increase in speed  in synchronized flow.
As predicted in~\cite{KernerBook,KernerBook2}, the stronger the over-acceleration, the more probable
should be the occurrence of the S$\rightarrow$F instability.

\subsection{Speed peak at downstream front of synchronized flow at on-ramp bottleneck  
  \label{Synch_KKSW_S}}

       \begin{figure} 
 \begin{center}
\includegraphics*[width=8 cm]{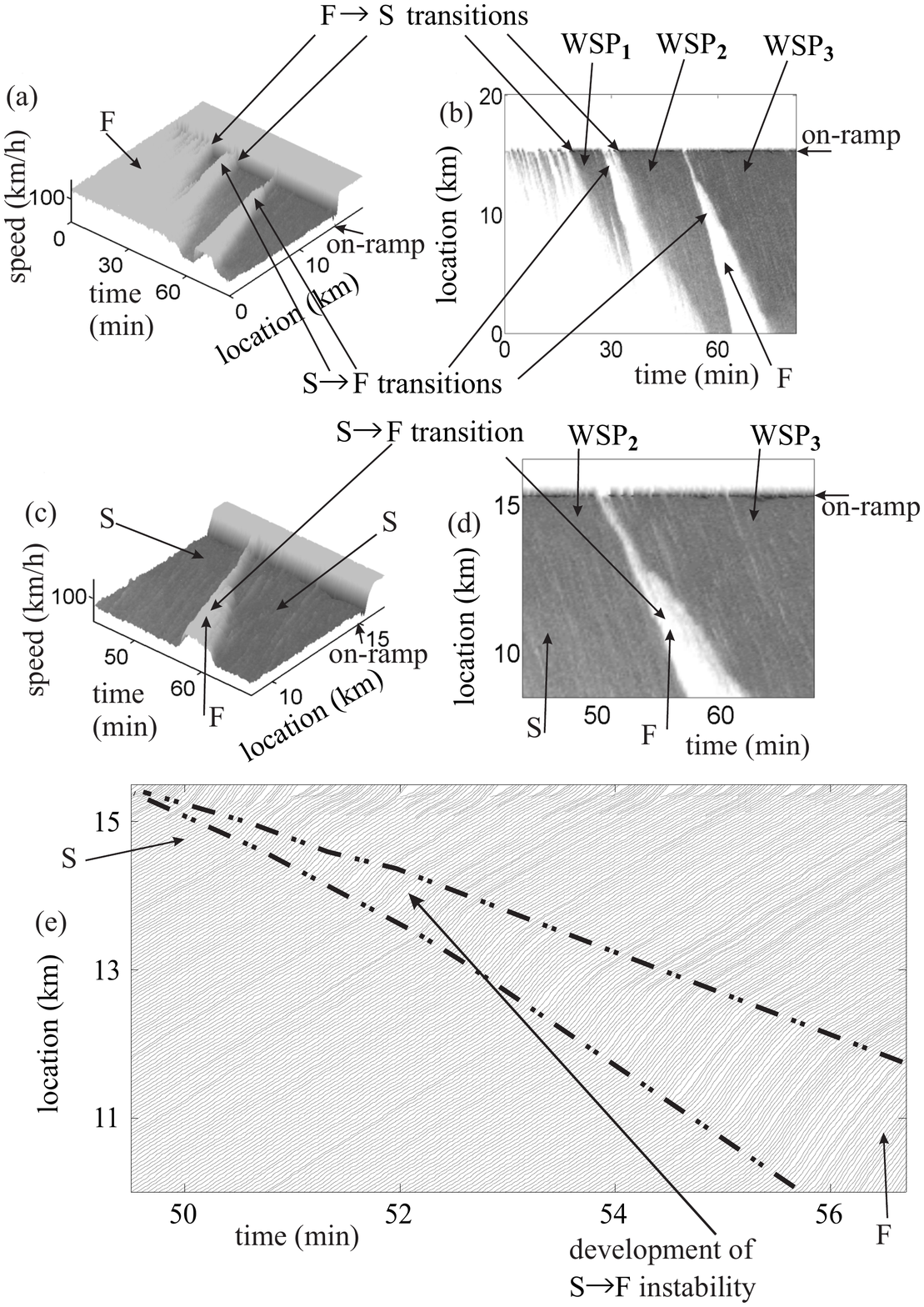}
 \end{center}
\caption{Simulations of the S$\rightarrow$F instability in synchronized flow leading to S$\rightarrow$F transition at on-ramp bottleneck on single-lane road with the KKSW CA model:
(a--d) Speed in space and time (a, c) and the same speed data
presented  by regions with variable shades of gray (in white regions the speed
is  equal to or higher than 110  km/h, in black regions the speed is zero) (b, d); figures (c, d) are, respectively, fragments of (a, b) in  larger scales in space and time. 
(e) Fragment of vehicle trajectories in space and time related to (c, d);
bold dashed-dotted curves in (e) mark the development of S$\rightarrow$F instability
in synchronized flow leading to S$\rightarrow$F transition.
F -- free flow, S -- synchronized flow, WSP -- widening synchronized flow pattern.
$q_{\rm on}=$ 360   vehicles/h,
$q_{\rm in}=$  1406   vehicles/h. 
On-ramp location $x_{\rm on}=$ 15 km. Merging region of the on-ramp is located within  15 km $\leq x \leq$ 15.3 km
(i.e., road locations within which vehicles can merge from the on-ramp lane onto the main road). Other model parameters are the same as those in Fig.~\ref{SteadyStates}.
 }
\label{SFS_Over_onramp_I_A}
\end{figure}

  \begin{figure} 
 \begin{center}
\includegraphics*[width=8 cm]{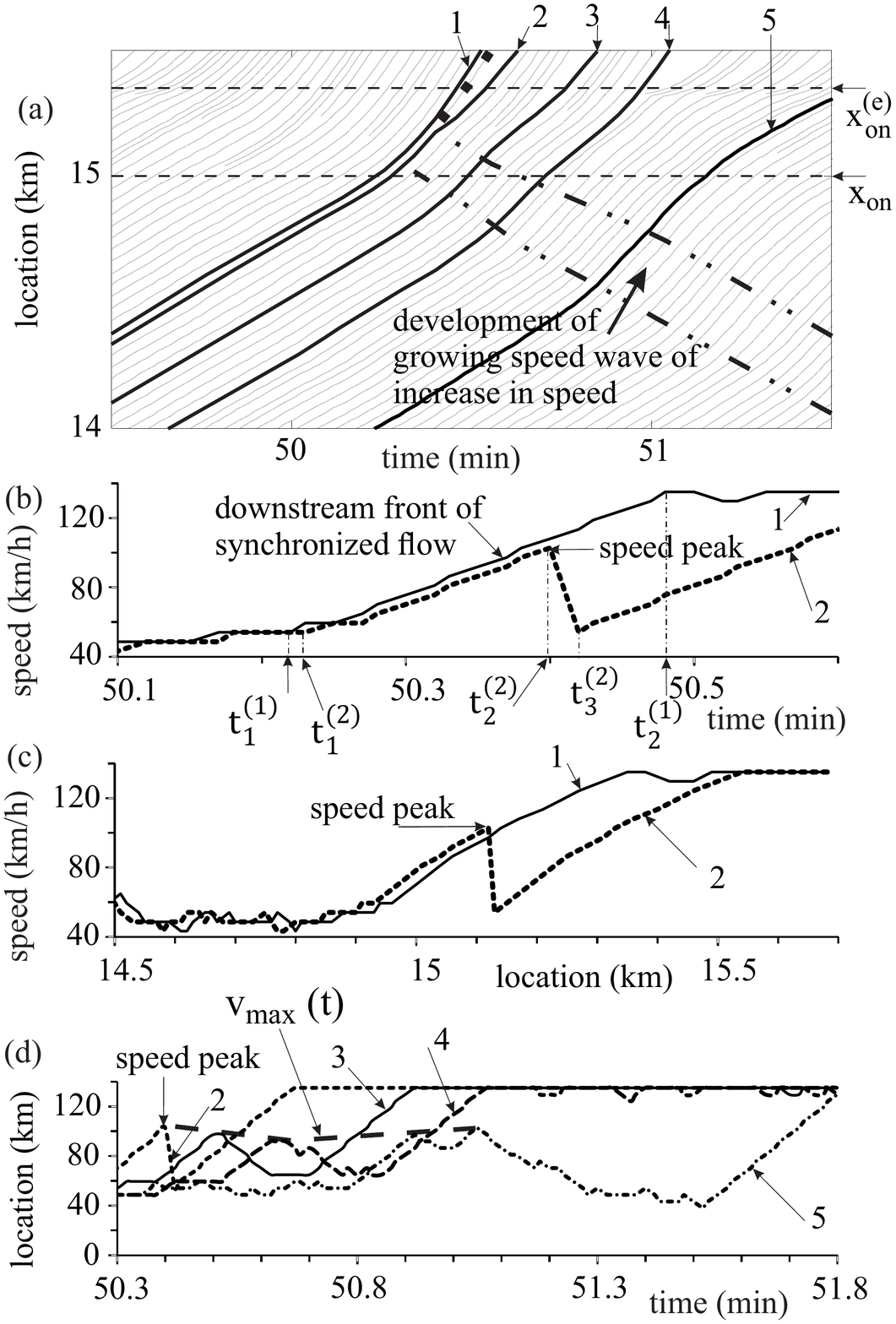}
 \end{center}
\caption{Speed peak at the downstream front of synchronized flow at bottleneck that
initiates S$\rightarrow$F instability     shown
in Fig.~\ref{SFS_Over_onramp_I_A} (c--e):
(a) Fragment of vehicle trajectories related to Fig.~\ref{SFS_Over_onramp_I_A} (e); bold dashed-dotted curves in (a) mark the  development of
the   speed wave of increase in speed within synchronized flow.
(b--d) Microscopic vehicle speed along trajectories as   time-functions (b, d) and road location functions (c).
In (b--d), vehicle trajectories are labeled by the same numbers as those in (a). $x_{\rm on}=$ 15 km and $x^{\rm (e)}_{\rm on}=15.3$ km
are, respectively, the beginning and the end of the merging region of the on-ramp which which vehicles can merge from the on-ramp onto the main road.
}
\label{SF_Over_onramp2_p_I_A}
\end{figure}

      \begin{figure} 
 \begin{center}
\includegraphics*[width=8 cm]{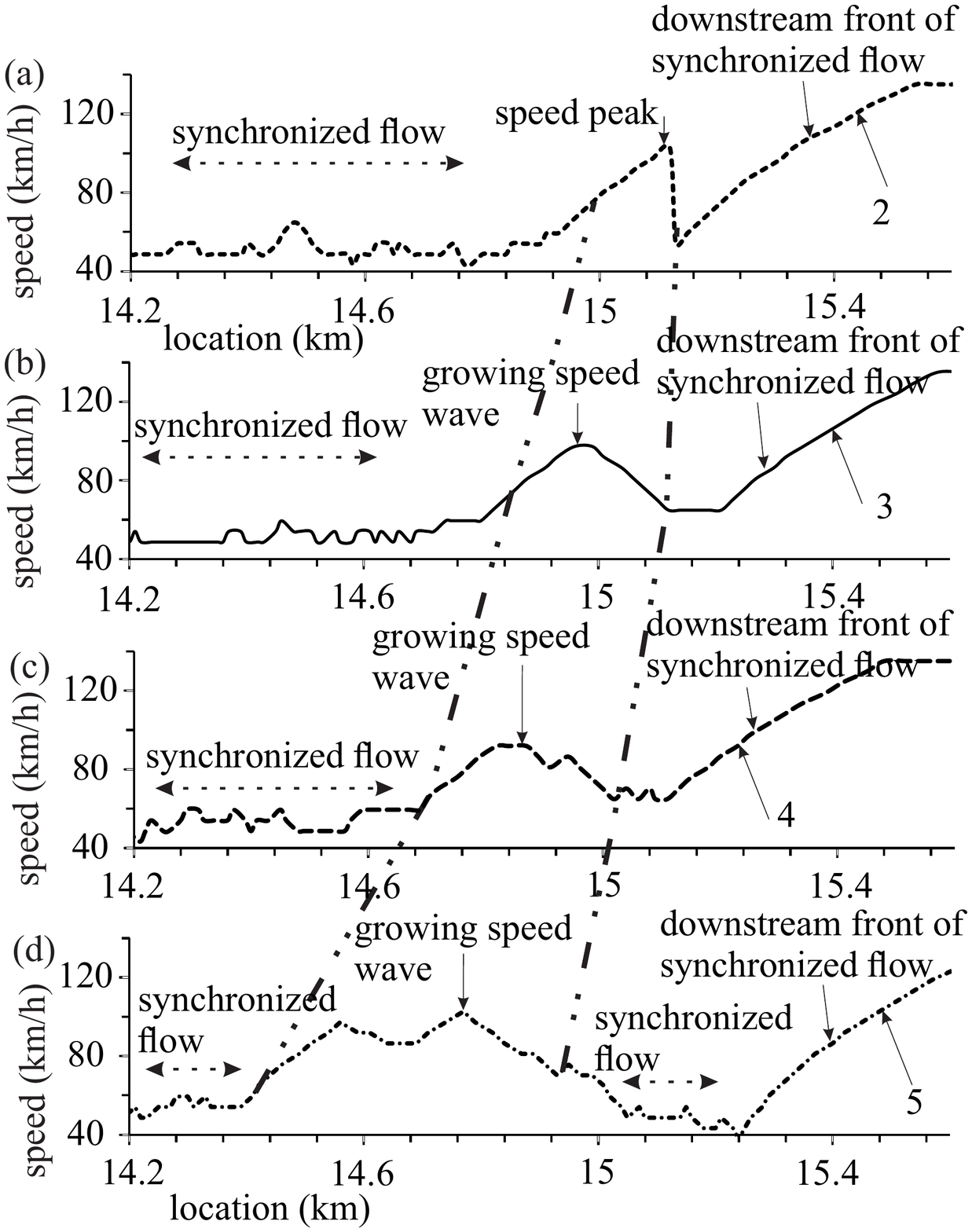}
 \end{center}
\caption{Transformation of speed peak shown in Fig.~\ref{SF_Over_onramp2_p_I_A} (c) into a
 growing speed wave of increase in speed propagating upstream within synchronized flow: 
(a--d) Microscopic vehicle speeds along trajectories as road location-functions
  labeled by the same numbers as those in Fig.~\ref{SF_Over_onramp2_p_I_A} (a).
 In (a--d), bold dashed-dotted curves mark
a growing speed wave of increase in speed within synchronized flow  as a function of road location.
}
\label{SF_Over_onramp2_1_1_I_A}
\end{figure}

   \begin{figure} 
 \begin{center}
\includegraphics*[width=8 cm]{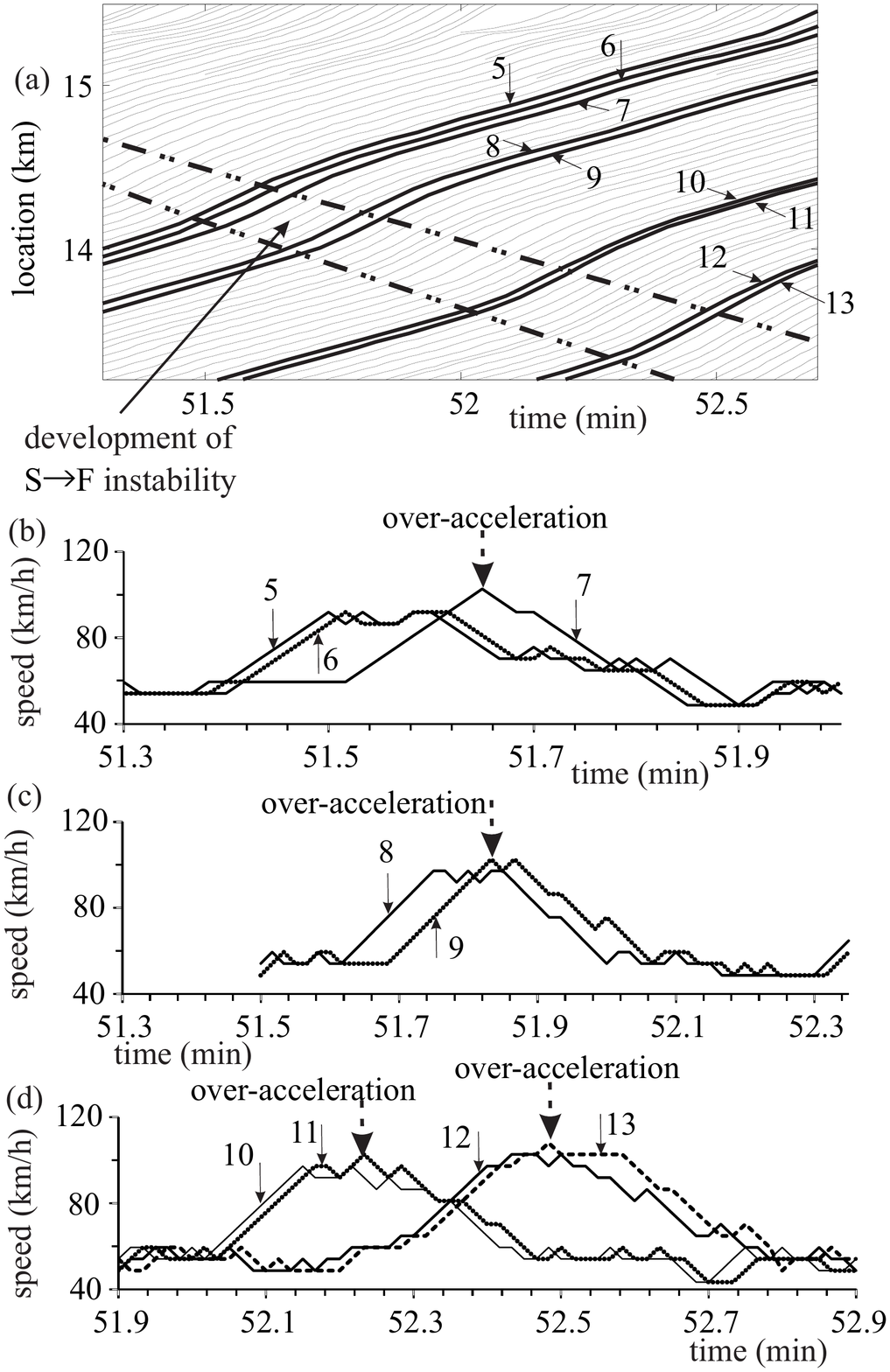}
 \end{center}
\caption{Effect of over-acceleration on   S$\rightarrow$F  instability   shown
in Fig.~\ref{SFS_Over_onramp_I_A} (c--e):
(a) Fragment of vehicle trajectories in space and time;  bold dashed-dotted curves mark
upstream propagation of the growing wave of increase in speed within synchronized flow.
(b--d) Microscopic vehicle speed along trajectories as time functions labeled by the same numbers as those in  (a).
}
\label{SF_Over_onramp2_2_I_A}
\end{figure}

  \begin{figure} 
 \begin{center}
\includegraphics*[width=8 cm]{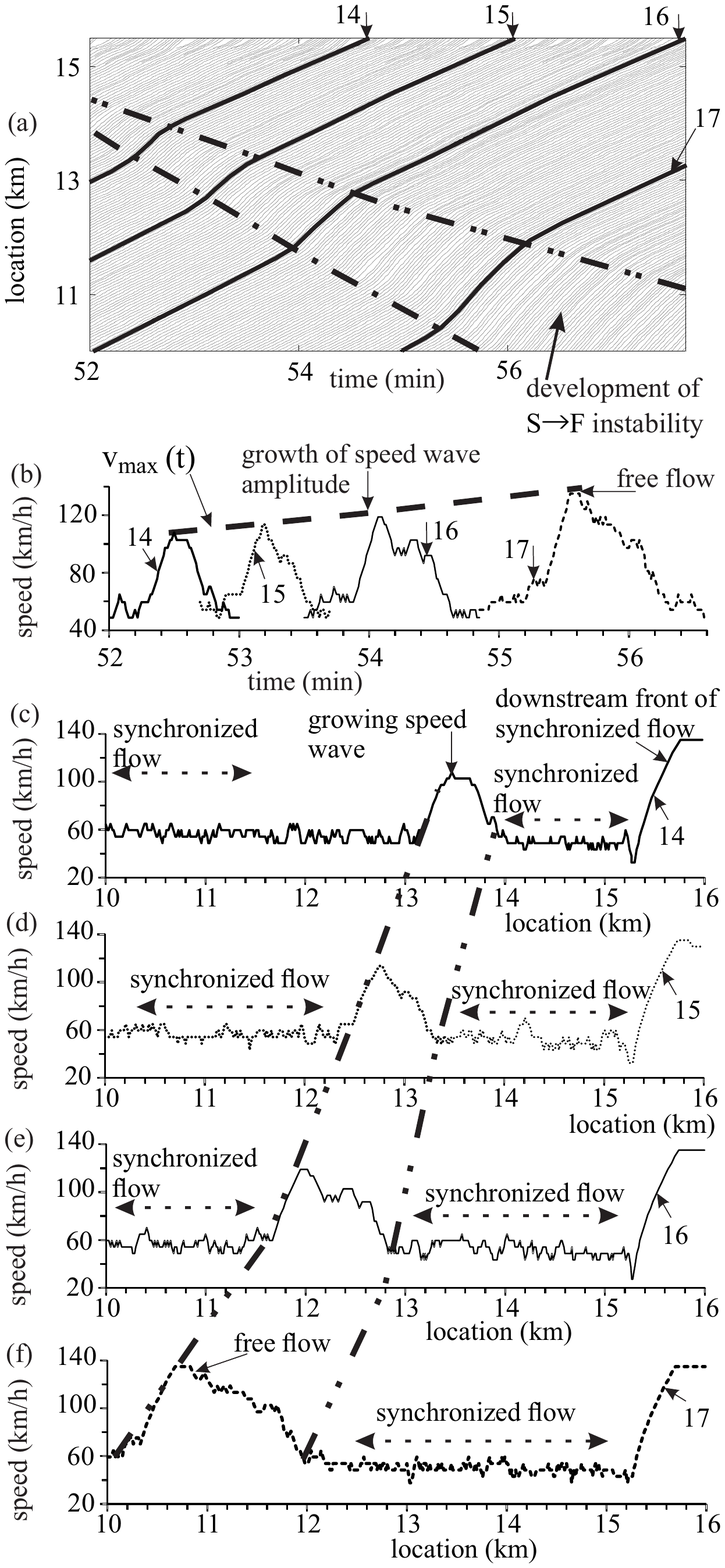}
 \end{center}
\caption{Subsequent development of S$\rightarrow$F instability, i.e.,
of growing speed wave of increase in speed within synchronized flow
shown in Fig.~\ref{SF_Over_onramp2_2_I_A} (a):
 (a) Fragment of vehicle trajectories in space and time;
 bold dashed-dotted curves in (a) mark
the  development of speed wave of increase in speed within synchronized flow.
 (b) Microscopic vehicle speeds along trajectories as time functions
  labeled by the same numbers as those in   (a); bold dashed curve marks
  the increase of the wave amplitude over time $v_{\rm max}(t)$.
   (c--f) Microscopic vehicle speeds along trajectories as road location-functions
  labeled by the same numbers as those in   (a);
  bold dashed-dotted curves mark
the development of growing speed wave of increase in speed within synchronized flow as a function of road location.
}
\label{SF_Over_onramp2_3_I_A}
\end{figure}

  In simulations of traffic flow on a single-lane road with an on-ramp bottleneck
  with the KKSW CA model, we find a sequence of F$\rightarrow$S and S$\rightarrow$F transitions at the bottleneck (labeled respectively by $\lq\lq$F$\rightarrow$S transitions"
and $\lq\lq$S$\rightarrow$F transitions" in  Fig.~\ref{SFS_Over_onramp_I_A} (a--d)).
  At chosen flow rates $q_{\rm on}$ and
$q_{\rm in}$ (Fig.~\ref{SFS_Over_onramp_I_A}), each of the F$\rightarrow$S  transitions leads  to the formation of a  widening synchronized flow pattern  (WSP)  
  at the bottleneck (labeled by $\lq\lq$$\rm WSP_{1}$", $\lq\lq$$\rm WSP_{2}$", and $\lq\lq$$\rm WSP_{3}$" in Fig.~\ref{SFS_Over_onramp_I_A}).
   To understand {\it microscopic features} of the S$\rightarrow$F instability, we   
  consider of an S$\rightarrow$F transition    shown in Fig.~\ref{SFS_Over_onramp_I_A} (c, d).
 
Microscopic features of the S$\rightarrow$F instability  (Fig.~\ref{SFS_Over_onramp_I_A} (e))
are as follows. 
Firstly,   a disturbance of increase in speed 
 emerges at the downstream front 
of synchronized flow at the on-ramp bottleneck (Fig.~\ref{SF_Over_onramp2_p_I_A}). We call this disturbance as $\lq\lq$speed peak"
  (labeled by $\lq\lq$speed peak" on trajectory 2  in Figs.~\ref{SF_Over_onramp2_p_I_A} (b--d)):
At time instant $t=t^{(1)}_{1}$  vehicle 1 begins to accelerate at the downstream front of synchronized flow 
 (Fig.~\ref{SF_Over_onramp2_p_I_A} (b, c)).
  Within the downstream front of synchronized flow, vehicle 1 accelerates continuously from a synchronized flow speed  
  to free flow downstream of the bottleneck. Vehicle 1 reaches a free flow speed at time instant $t^{(1)}_{2}$ (trajectory 1 in
  Fig.~\ref{SF_Over_onramp2_p_I_A} (b)). 
 A different situation is realized for vehicle 2 that follows vehicle 1 on the main road.
 
 After vehicle 1 has begun to accelerate,
   vehicle 2 begins also to accelerate at the downstream front of synchronized flow at    time instant
 $t^{(2)}_{1}$ (trajectory 2 in Fig.~\ref{SF_Over_onramp2_p_I_A} (b)). However,   a slower moving vehicle
 merges from on-ramp lane onto the main road between vehicles 1 and 2 (bold 
 dotted vehicle trajectory between vehicle trajectories 1 and 2  in Fig.~\ref{SF_Over_onramp2_p_I_A} (a)). 
 
Because vehicles 1 and 2 move on single-lane road, vehicle 2 cannot overtake the  vehicle merging from the on-ramp.
 As a result, vehicle 2 must   decelerate at time 
  $t^{(2)}_{2}$ (trajectory 2 in Fig.~\ref{SF_Over_onramp2_p_I_A} (b)).
  After the  vehicle merging from the on-ramp increases its speed considerably, vehicle 2
  can continue   acceleration to the free flow speed at time instant $t^{(2)}_{3}$ (trajectory 2 in Fig.~\ref{SF_Over_onramp2_p_I_A} (b)).
  This effect    
  leads to the occurrence of   a   speed peak 
  at the downstream front of synchronized flow   at the bottleneck (Fig.~\ref{SF_Over_onramp2_p_I_A} (b)).  
 
 \subsection{Over-acceleration effect as the 
 reason of  growing  speed wave of increase in speed within synchronized flow    
  \label{Wave_D_KKSW_S}} 

The    speed peak initiates a speed wave of increase in speed within synchronized flow. This speed wave propagates upstream.  This effect can be seen
in  Figs.~\ref{SF_Over_onramp2_p_I_A}(a, d) and~\ref{SF_Over_onramp2_1_1_I_A}. Firstly, while the wave propagates upstream, the maximum speed
$v_{\rm max}$ within the wave   does not  change considerably (Fig.~\ref{SF_Over_onramp2_p_I_A}(d)).
 
Later, the    speed wave begins to grow both in the amplitude and in the space 
 (Figs.~\ref{SF_Over_onramp2_1_1_I_A}--\ref{SF_Over_onramp2_3_I_A}). Finally, the growth of the wave leads to an S$\rightarrow$F transition
 at the bottleneck.
 The S$\rightarrow$F instability, i.e., the growth of the speed wave of
 a  local increase in  speed within synchronized flow is  caused by the over-acceleration effect. The growing speed wave of increase in speed
  in synchronized flow  can also be considered
  $\lq\lq$growing acceleration wave" in synchronized flow.
 To show the effect of over-acceleration on the  S$\rightarrow$F instability, we consider vehicle trajectories 5--13 within the growing  speed wave of increase in speed (Fig.~\ref{SF_Over_onramp2_2_I_A}).

The over-acceleration effect can be seen, if we compare the motion of vehicles 5, 6
with vehicle 7 that follow each other (Fig.~\ref{SF_Over_onramp2_2_I_A} (a)) within the speed wave of increase in   speed
(Fig.~\ref{SF_Over_onramp2_2_I_A} (b)).
Whereas vehicle 6 follows vehicle 5 without over-acceleration,
vehicle 7 accelerates while reaching the speed that exceeds
the speed of preceding vehicle 6 appreciably
(trajectories 6 and 7 in Fig.~\ref{SF_Over_onramp2_2_I_A} (b)). Although vehicle 6 begins to decelerate, nevertheless
  vehicle 7 accelerates. This acceleration of vehicle 7 occurs under conditions (\ref{gap_range}) and (\ref{Over_formula}). For this reason, the acceleration of vehicle 7 is an example
of the over-acceleration effect (labeled by $\lq\lq$over-acceleration" in Fig.~\ref{SF_Over_onramp2_2_I_A} (b)).

The effect of over-acceleration exhibits also vehicle 9 that follows vehicle 8,
vehicle 10 that follows vehicle 11
as well as vehicle 13 that follows vehicle 12 (trajectories 9--13 in Fig.~\ref{SF_Over_onramp2_2_I_A} (c ,d)).
The subsequent effects of over-acceleration
of different vehicles leads to the S$\rightarrow$F instability, i.e., to a growing wave of  the increase in speed within synchronized flow.
The speed wave grows both  in the amplitude and in the space extension
  during its upstream propagation within synchronized flow.
The subsequent development of this traffic flow instability caused by the over-acceleration
effect can be seen in Fig.~\ref{SF_Over_onramp2_3_I_A}.

\section{Nucleation Nature of S$\rightarrow$F   Instability    at Bottlenecks
  \label{Dis_Wave_KKSW_S}}
  
  \subsection{Random sequence of speed peaks at downstream front of synchronized flow  
  \label{Peaks_KKSW_S}} 
  
    \begin{figure} 
 \begin{center}
\includegraphics*[width=7.9 cm]{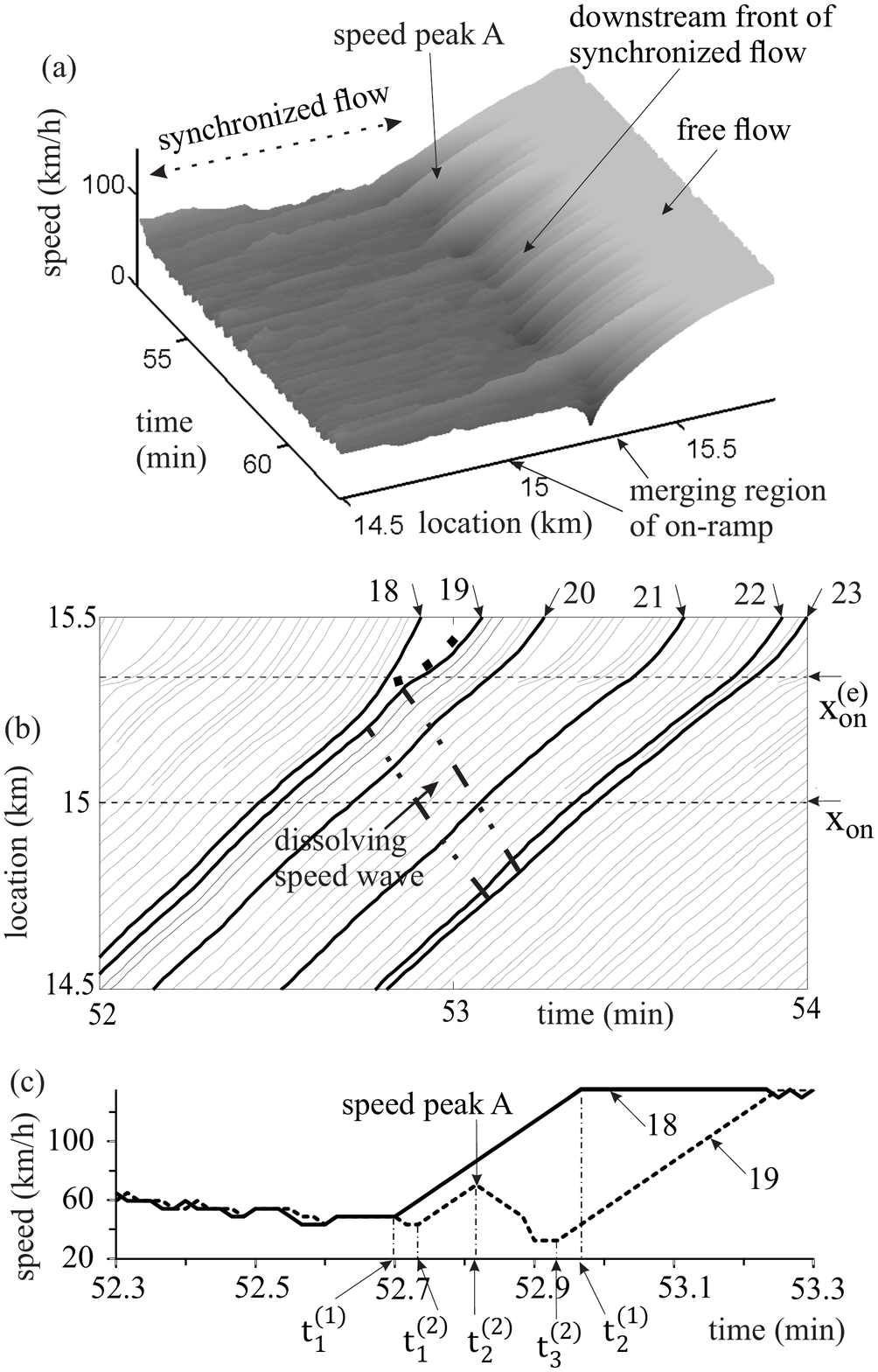}
 \end{center}
\caption{Speed peaks at   downstream front of synchronized flow at the on-ramp bottleneck: (a)
Speed in space and time; fragment of Fig.~\ref{SFS_Over_onramp_I_A} (a) for a time interval  that begins
after time instant at which the S$\rightarrow$F instability at bottleneck
has occurred; one of the speed peaks  in (a) is   marked by   $\lq\lq$speed peak A".
(b) Fragment of vehicle trajectories with a dissolving speed wave initiated by speed peak A in (a)
(the dissolving wave is marked by dotted-dashed curves). (c) Microscopic vehicle speed 
along trajectories as   time-functions showing the emergence of speed peak A; vehicle trajectories are labeled by the same numbers as those in
 (b). $x_{\rm on}=$ 15 km and $x^{\rm (e)}_{\rm on}=15.3$ km
are, respectively, the beginning and the end of the merging region of the on-ramp within which vehicles can merge from the on-ramp onto the main road.
}
\label{SFS_dis_onramp_I_A}
\end{figure}

    \begin{figure} 
 \begin{center}
\includegraphics*[width=8 cm]{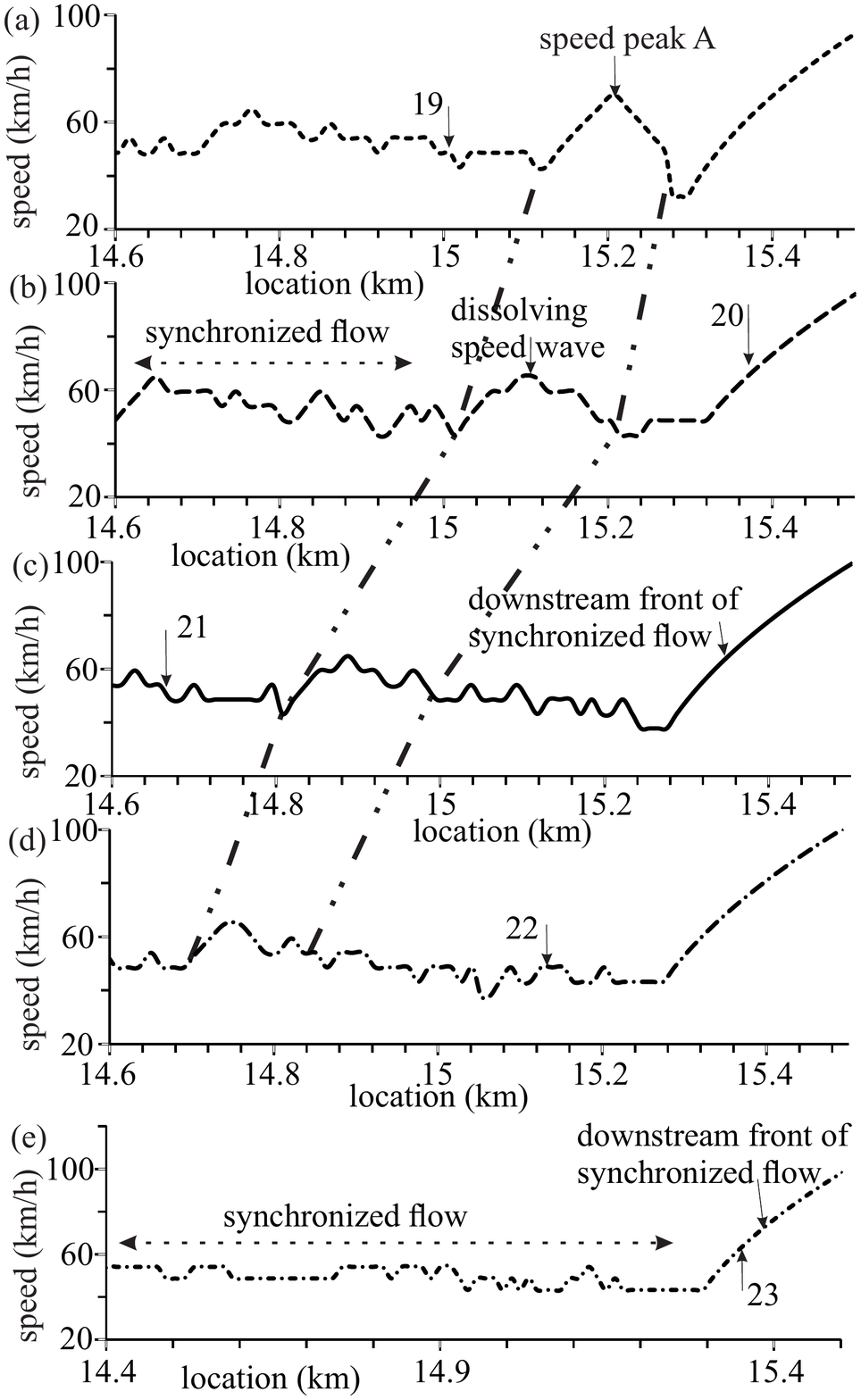}
 \end{center}
\caption{Dissolving speed wave within synchronized flow at the on-ramp bottleneck related to Fig.~\ref{SFS_dis_onramp_I_A} (b, c):
(a--e) Microscopic speed along vehicle trajectories as road location functions. Vehicle trajectories are labeled by the same numbers as those in
Fig.~\ref{SFS_dis_onramp_I_A} (b). Bold dashed-dotted curves mark the propagation of dissolving speed wave in space.
}
\label{SFS_dis_onramp2_I_A}
\end{figure}

There can be many speed peaks that occur randomly     at the downstream front of synchronized flow at the on-ramp bottleneck (Fig.~\ref{SFS_dis_onramp_I_A} (a)). 
The physics of all speed peaks 
shown in Fig.~\ref{SFS_dis_onramp_I_A} is the same as   discussed above (Sec.~\ref{Synch_KKSW_S}). 

As an example, we consider
a speed peak labeled by $\lq\lq$speed peak A" in Fig.~\ref{SFS_dis_onramp_I_A} (a).
Due to slow vehicle merging from the on-ramp onto the main road (bold dotted
vehicle trajectory between vehicle trajectories 18 and 19 in Fig.~\ref{SFS_dis_onramp_I_A} (b)),  
vehicle 19 moving on the main road at time instant $t^{(2)}_{2}$  should change      acceleration at the downstream front of synchronized flow to deceleration 
(Fig.~\ref{SFS_dis_onramp_I_A} (c)); other time instants marked in Fig.~\ref{SFS_dis_onramp_I_A} (c)
have also the same sense as those in Fig.~\ref{SF_Over_onramp2_p_I_A} (b).
As a result of this  deceleration of vehicle 19,
 speed peak A emerges (Fig.~\ref{SFS_dis_onramp_I_A} (a, c)).

 \subsection{Dissolving speed wave of increase in speed within synchronized flow  at bottleneck
  \label{Diss_KKSW_S}}
 
Speed peak A initiates a speed wave of increase in speed within synchronized flow
that propagates upstream. However, rather than an S$\rightarrow$F instability occurs  discussed in Secs.~\ref{Synch_KKSW_S} and~\ref{Wave_D_KKSW_S},
the wave is fully dissolved about 0.3 km
upstream of the beginning of the on-ramp merging region at $x=$ 15 km. We call this wave as $\lq\lq$dissolving speed wave" of increase in speed
in synchronized flow
(Figs.~\ref{SFS_dis_onramp_I_A} (b)  and~\ref{SFS_dis_onramp2_I_A} (e)).

The speed peak shown in Fig.~\ref{SF_Over_onramp2_p_I_A}, which
  initiates the S$\rightarrow$F instability (Secs.~\ref{Synch_KKSW_S} and~\ref{Wave_D_KKSW_S}),
  and speed peak   that does not initiate  an S$\rightarrow$F instability  differ in their amplitudes:
  The speed within the peak shown in Fig.~\ref{SF_Over_onramp2_p_I_A} is about 98 km/h; 
  the speed within peak A is considerably smaller (about 70   km/h).
  All other speed peaks that emerge
  at the downstream front of synchronized flow (Fig.~\ref{SFS_dis_onramp_I_A} (a))
  exhibit also considerably smaller amplitudes than that
   of the speed peak shown in Fig.~\ref{SF_Over_onramp2_p_I_A}. As  a result, all waves of increase in speed   within synchronized flow
  that
  the other speed peaks   initiate are dissolving speed waves. A dissolving speed wave of increase in speed
  in synchronized flow  can also be considered
  $\lq\lq$dissolving acceleration wave" in synchronized flow.
  
  We have found that if the speed peak amplitude is equal to or larger than some critical one, the speed peak is a nucleus
  for an S$\rightarrow$F instability (Secs.~\ref{Synch_KKSW_S} and~\ref{Wave_D_KKSW_S}). Contrarily,
  if the peak amplitude is smaller than the critical one (as this is the case for all speed peaks in Fig.~\ref{SFS_dis_onramp_I_A} (a)), 
  the speed peak is smaller than a nucleus for an S$\rightarrow$F instability: Instead of
  the S$\rightarrow$F instability, the peak
  initiates a dissolving wave of the increase in speed within synchronized flow (Figs.~\ref{SFS_dis_onramp_I_A} (b)  and~\ref{SFS_dis_onramp2_I_A}).
  
  The physics of the nucleation nature of an S$\rightarrow$F instability is as follows.
  The over-acceleration effect is able to overcome   speed adaptation between following each other vehicles (speed adaptation effect)
  only if the speed within the speed wave is large enough: When the  over-acceleration effect is stronger than the speed adaptation
  effect within the speed wave,  as that occurs in Fig.~\ref{SF_Over_onramp2_2_I_A}, the S$\rightarrow$F instability is realized.
  Otherwise, when during the speed wave propagation the speed adaptation effect
  suppresses the over-acceleration within synchronized flow,
  the speed wave dissolves over time, i.e., no S$\rightarrow$F instability is realized (Fig.~\ref{SFS_dis_onramp2_I_A}(b--e)).

 \section{Random time-delayed traffic  breakdown as   result of S$\rightarrow$F instability        \label{Nuc_Sec}}

    As already found in~\cite{KKl,KKW,KKl2003A}, there is a random time delay $T^{\rm (B)}$ between
   the beginning of a simulation realization and the time instant at which traffic breakdown (F$\rightarrow$S transition) occurs  resulting
    in the emergence of   a congested pattern at the bottleneck. At chosen flow rates $q_{\rm on}$ and
$q_{\rm in}$,   the congested pattern is   an WSP (Fig.~\ref{FS_delay_onramp_A}).

   The microscopic nature of a random time delay of traffic breakdown at the bottleneck revealed below allows us
   to understand that and how an S$\rightarrow$F instability governs traffic breakdown.  
However, before we should
   understand microscopic features of traffic breakdown    at the bottleneck (Sec.~\ref{Mic_FS_S}).
 
 \subsection{Microscopic features of traffic breakdown  (F$\rightarrow$S  transition) at bottleneck  
  \label{Mic_FS_S}}

          \begin{figure} 
 \begin{center}
\includegraphics*[width=7.8 cm]{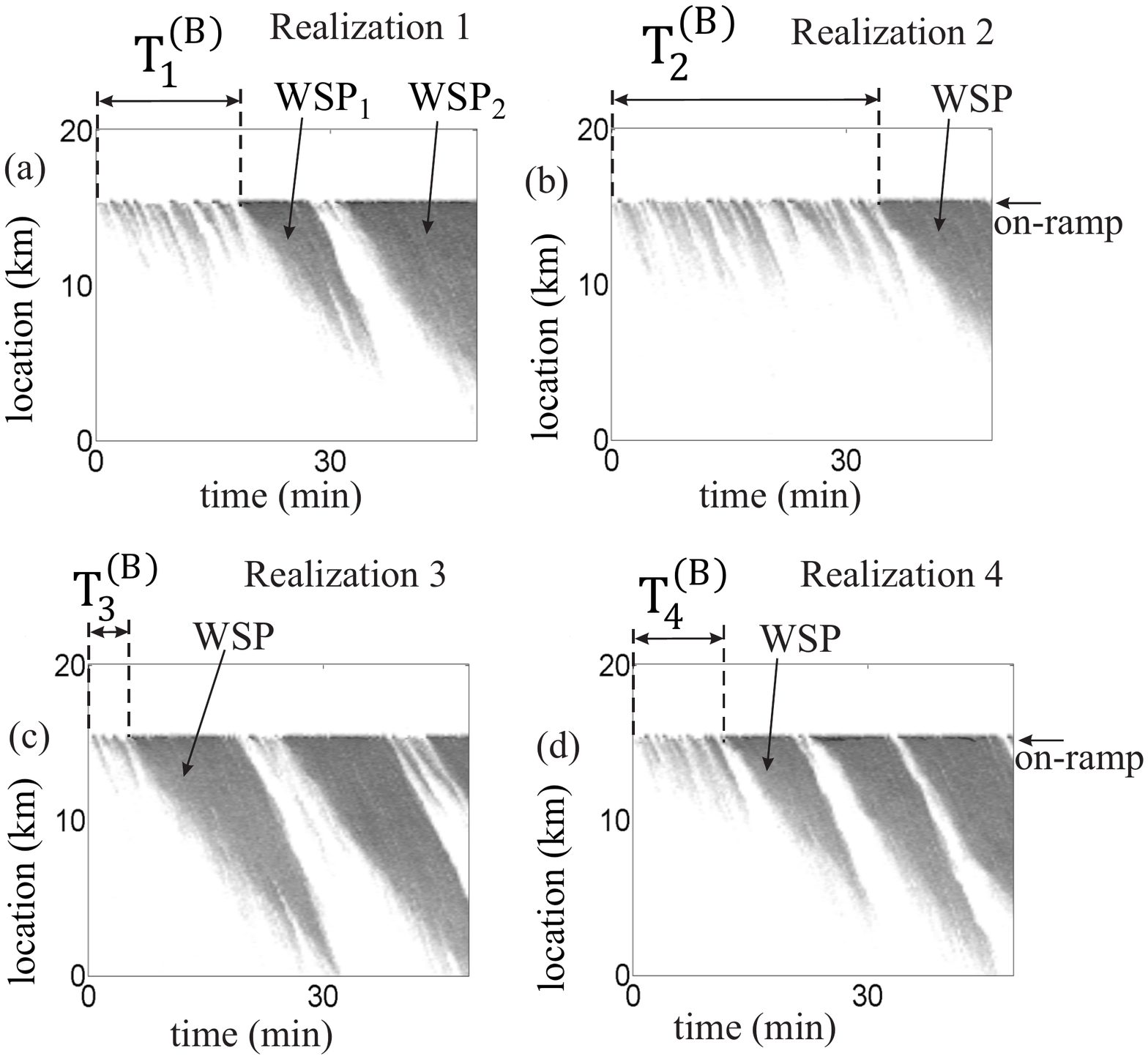}
 \end{center}
\caption{Random time delay of traffic breakdown (F$\rightarrow$S transition) at on-ramp bottleneck:
(a--d)   Speed in space and time  for four different simulation realizations (runs)
presented  by regions with variable shades of gray (in white regions the speed
is  equal to or higher than 110  km/h, in black regions the speed is zero).  Different   realizations  are made at the same   model parameters, however,
   at different   initial    values $r$ in formulae (\ref{rand_p1}) and (\ref{rand_p})
   at time instant $t=0$. Realization 1 in   (a) is a fragment
of Fig.~\ref{SFS_Over_onramp_I_A} (b), i.e., realization 1 is  the simulation realization studied in
  Figs.~\ref{SFS_Over_onramp_I_A}--\ref{SFS_dis_onramp2_I_A}. Time delays of traffic breakdown $T^{\rm (B)}$ in different simulation realizations 1--4 are
$T^{\rm (B)}_{1}=$ 19 min (a), $T^{\rm (B)}_{2}=$35 min (b),  $T^{\rm (B)}_{3}=$7 min (c), and $T^{\rm (B)}_{4}=$13 min (d). $q_{\rm on}=$ 360   vehicles/h,
$q_{\rm in}=$  1406   vehicles/h.
}
\label{FS_delay_onramp_A} 
\end{figure}

       \begin{figure} 
 \begin{center}
\includegraphics*[width=8 cm]{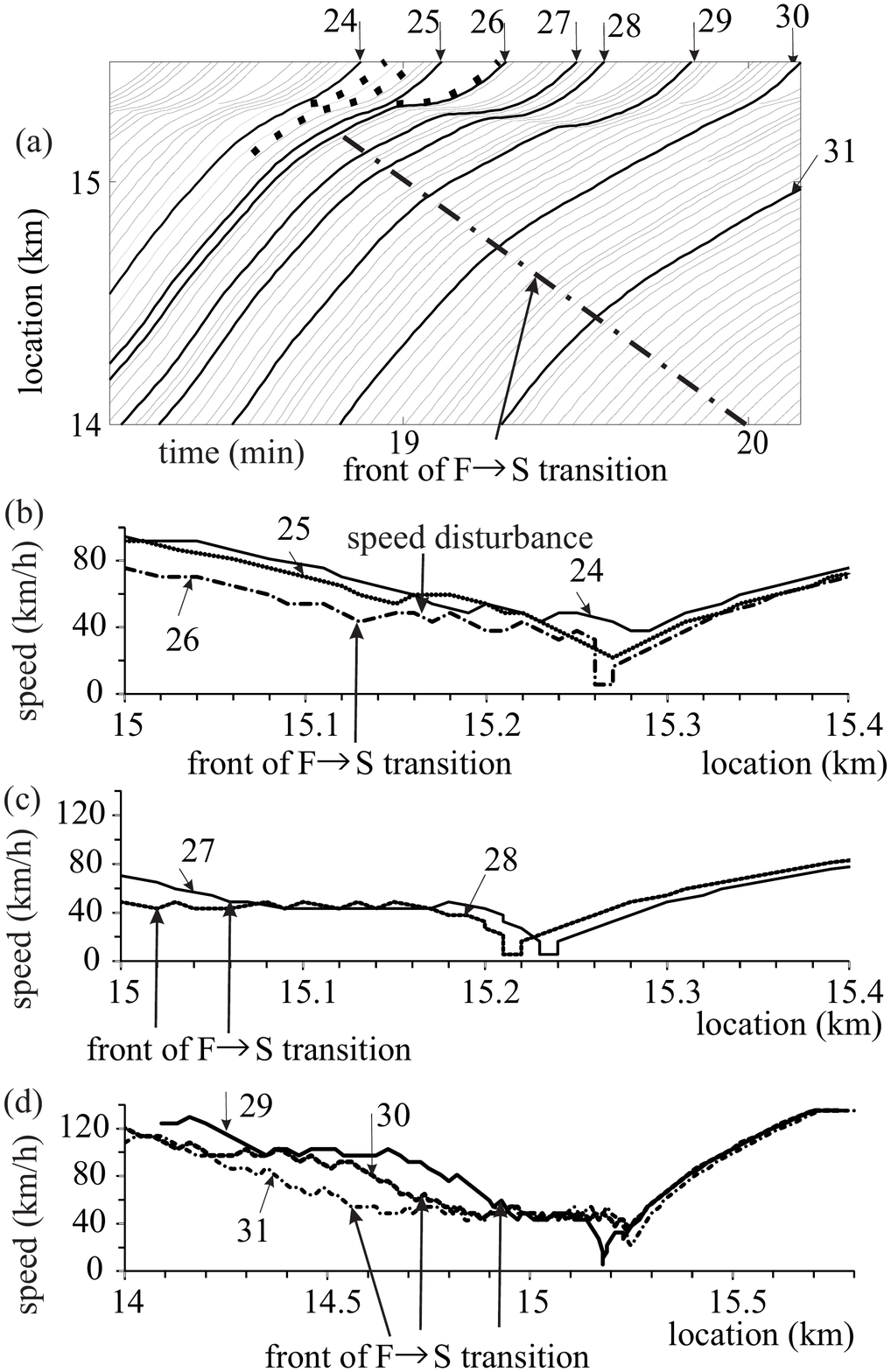}
 \end{center}
\caption{Traffic breakdown (F$\rightarrow$S transition) at on-ramp bottleneck that leads to the formation of ${\rm WSP_{1}}$ in Fig.~\ref{FS_delay_onramp_A} (a):
(a) Fragments of vehicle trajectories related to Fig.~\ref{FS_delay_onramp_A} (a); bold dashed-dotted curve marks the propagation of the upstream front of synchronized flow
(labeled by $\lq\lq$front of F$\rightarrow$S transition")  in space and time.
(b--d) Microscopic speed along vehicle trajectories as road location functions. Vehicle trajectories are labeled by the same numbers as those in
(a).
}
\label{FS_Over_onrampLoc_A}
\end{figure}

We have found that in each of the simulation realizations (Fig.~\ref{FS_delay_onramp_A}), traffic breakdown (F$\rightarrow$S transition) exhibits   the following common   microscopic features: 

(i)
Vehicles    that merge onto the main road from the on-ramp (vehicle trajectories labeled by bold dotted curves in 
Fig.~\ref{FS_Over_onrampLoc_A} (a)) force
vehicles moving on the main road to decelerate strongly. This results in the formation of a speed disturbance of decrease in speed.
The upstream front of the disturbance
begins to propagate upstream of the bottleneck  (labeled by $\lq\lq$speed disturbance" on
vehicle trajectory 26 in Fig.~\ref{FS_Over_onrampLoc_A} (b)).

(ii) Due to speed adaptation of vehicles following this decelerating vehicle on the main road (vehicle trajectories
 27--31 in Fig.~\ref{FS_Over_onrampLoc_A} (a, c, d)),
synchronized flow region appears that upstream front propagates 
upstream (labeled by $\lq\lq$front of F$\rightarrow$S transition" in Fig.~\ref{FS_Over_onrampLoc_A}).

(iii) After traffic breakdown has occurred,   many speed peaks appear in the synchronized flow at the bottleneck (not shown). The microscopic features
of these peaks are qualitatively the same as those shown in Fig.~\ref{SFS_dis_onramp_I_A} (a, c). In particular,  the speed peaks lead to formation of a
 speed wave of increase in speed that propagates upstream within the synchronized flow.
 During a long enough time interval (time interval of the existence of  $\rm WSP_{1}$ shown in Fig.~\ref{FS_delay_onramp_A} (a)), all speed waves are dissolving ones.  The dissolving speed waves (not shown)
   exhibit the same microscopic features   as those shown in Figs.~\ref{SFS_dis_onramp_I_A} (b) and~\ref{SFS_dis_onramp2_I_A}.

   \subsection{Microscopic features of sequence of F$\rightarrow$S$\rightarrow$F transitions at bottleneck  
  \label{Mic_FSF_S}}  
  
            \begin{figure} 
 \begin{center}
\includegraphics*[width=8 cm]{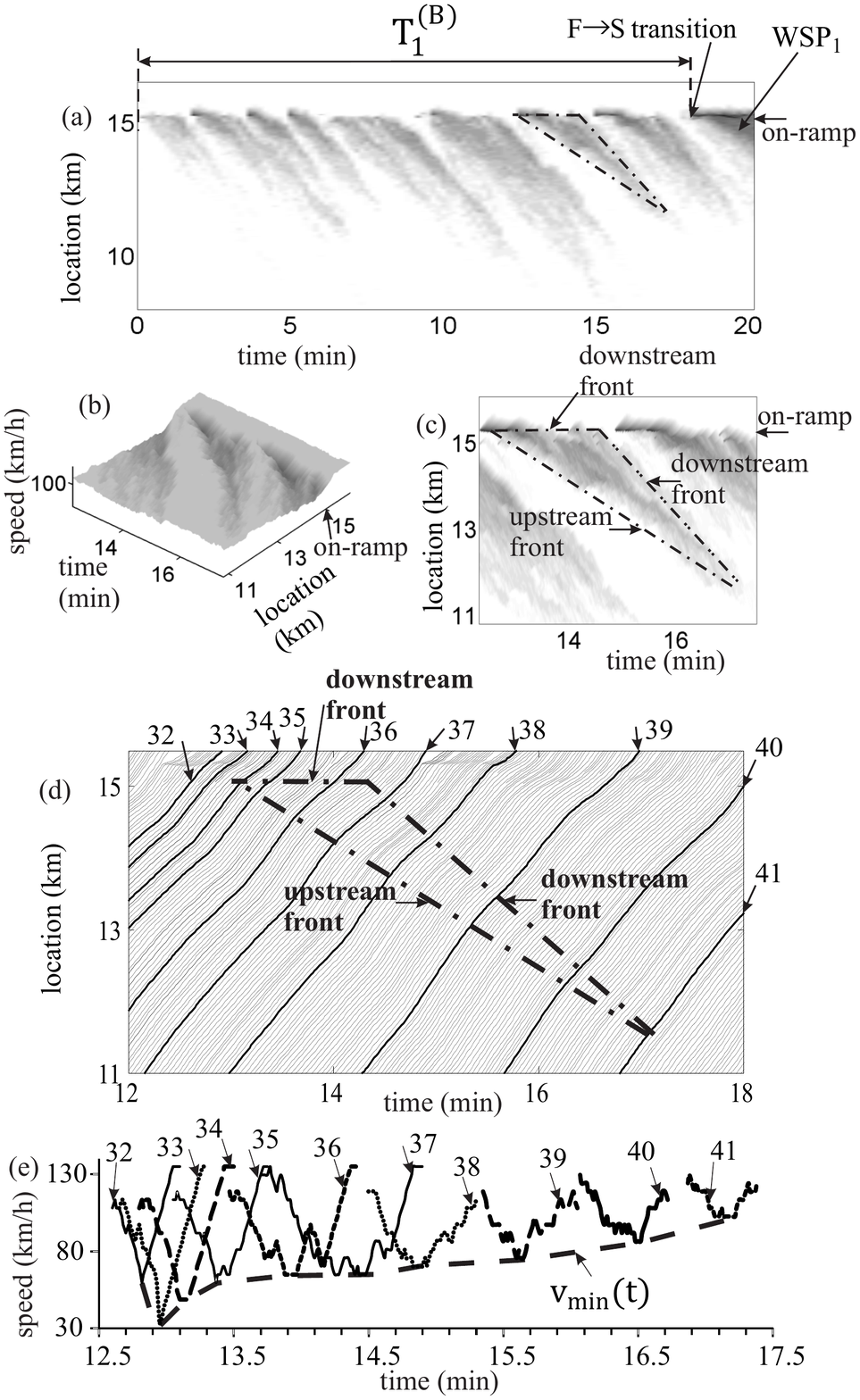}
 \end{center}
\caption{Simulations of F$\rightarrow$S$\rightarrow$F transitions within a permanent speed disturbance at on-ramp bottleneck: 
(a) Speed in space and time  
presented  by regions with variable shades of gray (in white regions the speed
is  equal to  or higher than 100  km/h, in black regions the speed is equal to 20 km/h)   within time delay of traffic breakdown related to Fig.~\ref{FS_delay_onramp_A} (a).
(b, c) Speed in space and time (b) and the same speed data
presented  by regions with variable shades of gray (c) for  a short time interval in  (a). 
(d) Fragment of vehicle trajectories in space and time related to (b, c). (e) Microscopic vehicle speeds along trajectories as time functions
labeled by the same numbers as those in (d).
}
\label{FSF_KKSW_traj1_F_A}
\end{figure}

  \begin{figure} 
 \begin{center}
\includegraphics*[width=7.8 cm]{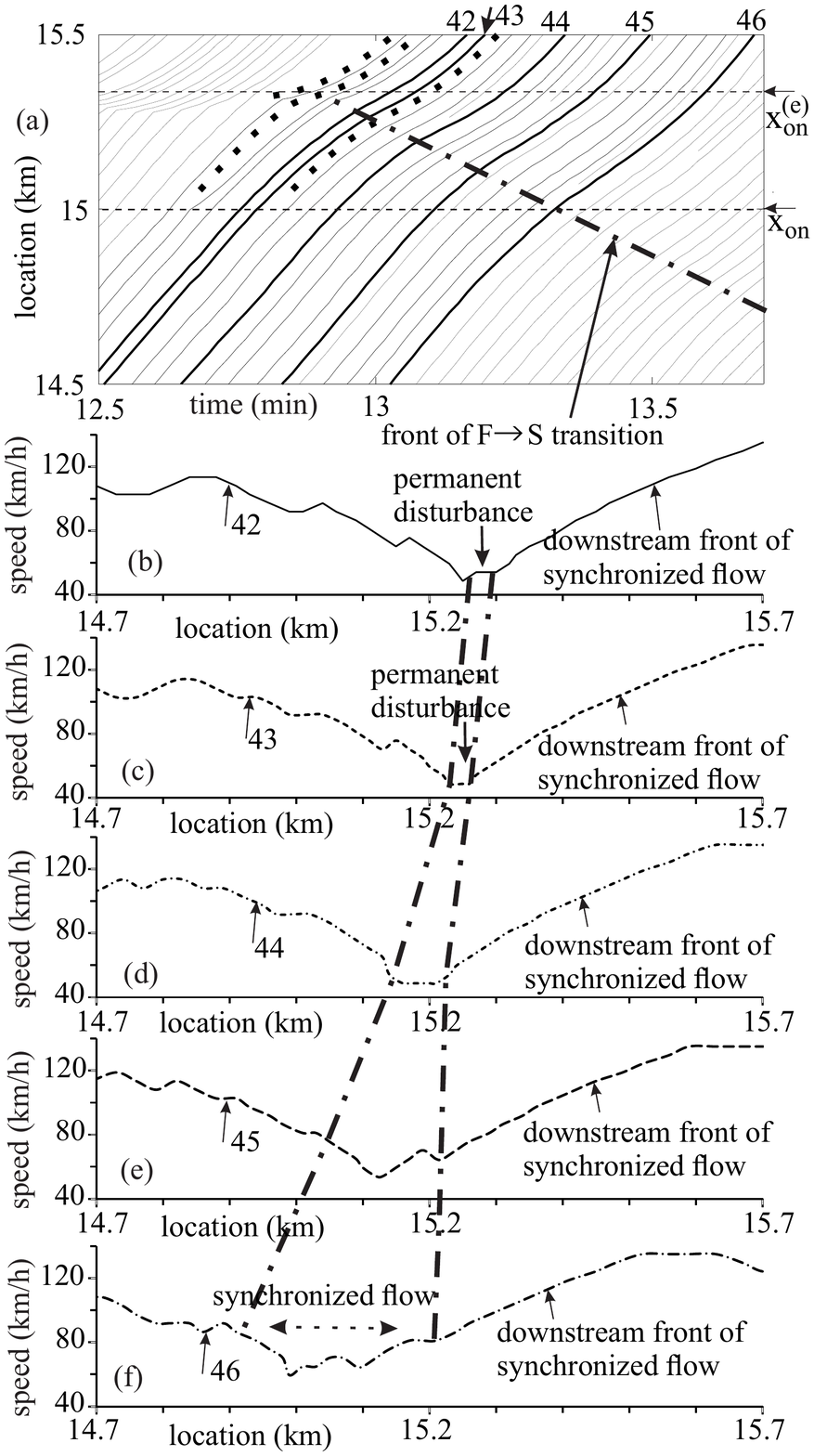}
 \end{center}
\caption{Simulations of  an F$\rightarrow$S  transition    within a permanent speed disturbance (labeled by $\lq\lq$permanent disturbance") at on-ramp bottleneck:
(a) Fragment of vehicle trajectories in space and time related to Fig.~\ref{FSF_KKSW_traj1_F_A} (b, c).  
 (b--f) Microscopic vehicle speeds along trajectories as  
road location-functions  
labeled by the same numbers as those in   (a).
}
\label{FSF_KKSW_traj2_F_A}  
\end{figure}

  \begin{figure} 
 \begin{center}
\includegraphics*[width=7.5 cm]{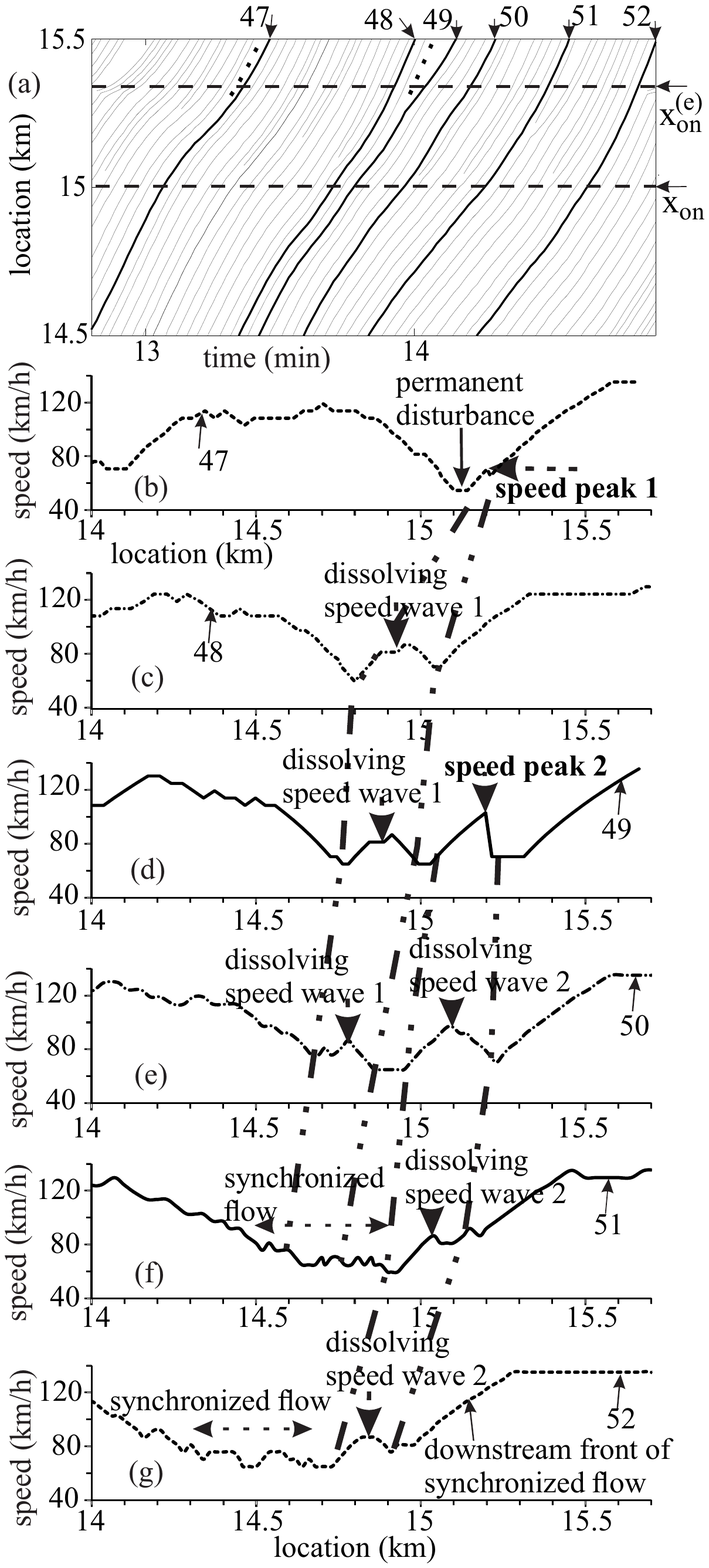}
 \end{center}
\caption{Simulations of speed peaks 1 and 2 with resulting dissolving speed waves 1 and 2 of increase in speed within synchronized flow at on-ramp bottleneck:
(a) Fragment of vehicle trajectories in space and time related to Fig.~\ref{FSF_KKSW_traj1_F_A} (b, c).   
(b--g)  Microscopic vehicle speeds along trajectories as  
road location-functions  
labeled by the same numbers as those in   (a).
}
\label{FSF_KKSW_traj3_F_A}
\end{figure}

  \begin{figure} 
 \begin{center}
\includegraphics*[width=8 cm]{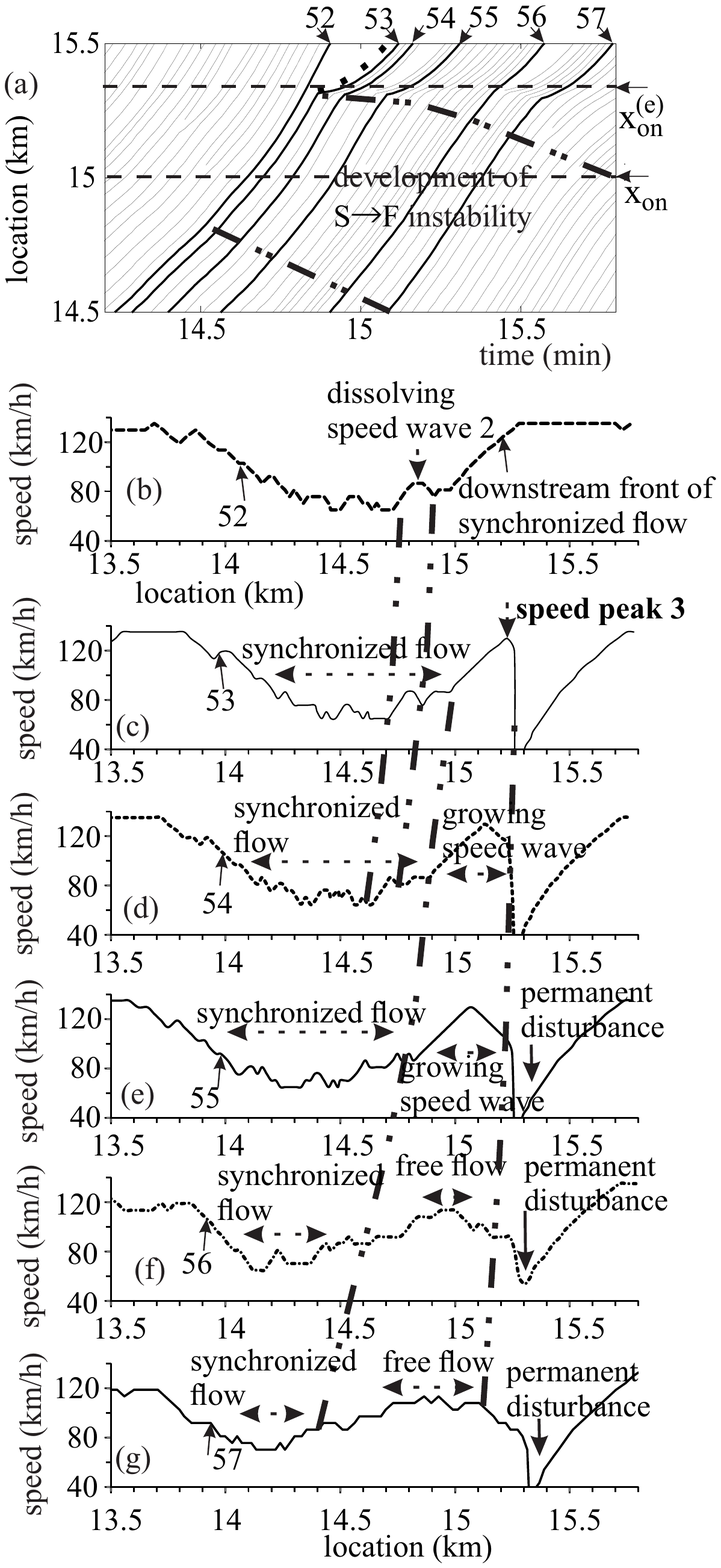}
 \end{center}
\caption{Simulations of a growing speed wave of increase in speed within synchronized flow at on-ramp bottleneck (region bounded by bold dashed-dotted curves
labeled by $\lq\lq$growing speed wave"):
(a) Fragment of vehicle trajectories in space and time related to Fig.~\ref{FSF_KKSW_traj1_F_A} (b, c).  
 (b--g) Microscopic vehicle speeds along trajectories as  
road location-functions  
labeled by the same numbers as those in (a). A dissolving
speed wave (region bounded by bold dashed-dotted curves labeled by $\lq\lq$dissolving speed wave 2")
is a continuation of the   dissolving speed wave 2 shown in Fig.~\ref{FSF_KKSW_traj3_F_A} (e--g).
}
\label{FSF_KKSW_traj4_F_A}
\end{figure}

    \begin{figure} 
 \begin{center}
\includegraphics*[width=8 cm]{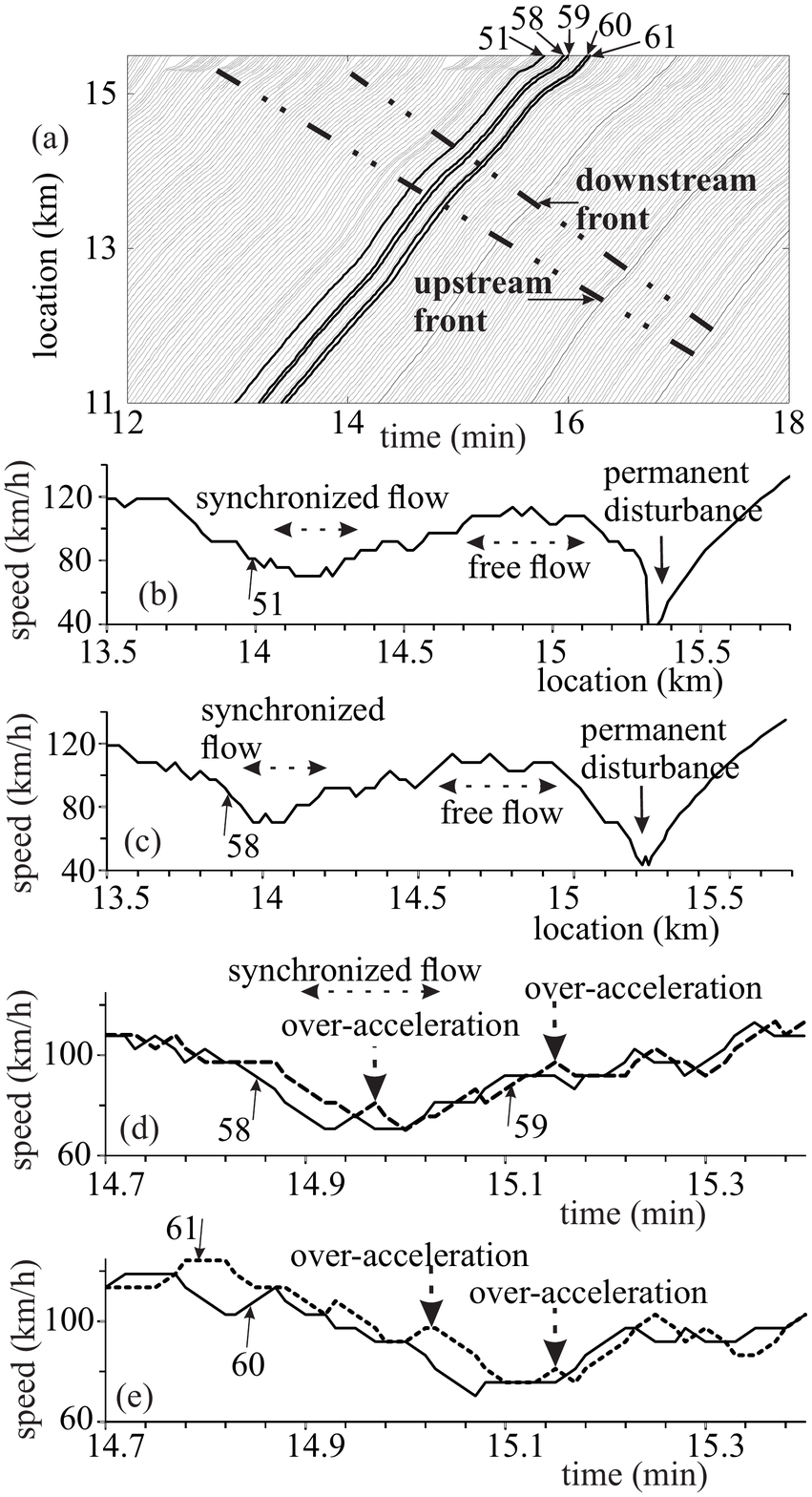} 
 \end{center}
\caption{Simulations of over-acceleration effect that leads to S$\rightarrow$F instability, i.e., to the
growing speed wave of increase in speed within synchronized flow at on-ramp bottleneck:  
(a) Fragment of vehicle trajectories in space and time related to Fig.~\ref{FSF_KKSW_traj1_F_A}  (b--d).  
 (b--e) Microscopic vehicle speeds along trajectories as  
road location-functions  (b, c) and time-functions (d, e)
labeled by the same numbers as those in (a).
}
\label{FSF_KKSW_traj7_F_A}
\end{figure}

 We have found that during the time delay of traffic breakdown $0<t<T^{\rm (B)}_{1}$ (Figs.~\ref{FS_delay_onramp_A} (a) and~\ref{FSF_KKSW_traj1_F_A} (a))
   there is
  a   permanent spatiotemporal competition  between the speed adaptation effect supporting an
  F$\rightarrow$S transition
   and the over-acceleration effect supporting    an S$\rightarrow$F instability that counteracts   the emergence of synchronized flow.
   This competition results in the occurrence of a permanent speed decrease in a neighborhood of the bottleneck that we call $\lq\lq$permanent
   speed disturbance" at the bottleneck. There can be distinguished two   cases of this competition: 
   
   (i)   There is a noticeable time lag
   between the beginning of an F$\rightarrow$S transition due to the speed adaptation   
   and the beginning of an S$\rightarrow$F instability due to over-acceleration that prevents the formation of a congested pattern   at the bottleneck; this case we call 
   $\lq\lq$a sequence of F$\rightarrow$S$\rightarrow$F transitions" at the bottleneck. 
   
   (ii) There is a 
   spatiotemporal $\lq\lq$overlapping" of the speed adaptation and over-acceleration effects  
  (Sec.~\ref{Mic_Perm_Dis_S}).

One of the   sequences of F$\rightarrow$S$\rightarrow$F transitions within the permanent speed disturbance at on-ramp bottleneck is marked
by dashed-dotted curves in Fig.~\ref{FSF_KKSW_traj1_F_A}   (a, c). An F$\rightarrow$S transition and a return S$\rightarrow$F  
 transition that build the sequence  of F$\rightarrow$S$\rightarrow$F transitions  
 are explained as follows (Figs.~\ref{FSF_KKSW_traj1_F_A}    (b--e)--\ref{FSF_KKSW_traj7_F_A}).

\subsubsection{F$\rightarrow$S  transition}  
 
  After several slow moving vehicles have merged from the on-ramp onto the main road (bold dotted vehicle trajectories   in Fig.~\ref{FSF_KKSW_traj2_F_A}    (a)),
  the following vehicles on the main road have to decelerate strongly due to the speed adaptation effect (vehicle trajectories 42 and 43  in Fig.~\ref{FSF_KKSW_traj2_F_A}    (a--c)).
  This results in the upstream propagation of synchronized flow upstream of the bottleneck, i.e.,
  an  F$\rightarrow$S  transition occurs (vehicle trajectories 42--46  in Fig.~\ref{FSF_KKSW_traj2_F_A}    (a--f)).
  Microscopic features of this F$\rightarrow$S  transition
  (in particular,  the upstream propagation of the upstream front of synchronized flow labeled by $\lq\lq$front of F$\rightarrow$S  transition"
  in Fig.~\ref{FSF_KKSW_traj1_F_A}    (a)) are qualitatively the same as those shown in Fig.~\ref{FS_Over_onrampLoc_A}.
  
  Moreover,  after the F$\rightarrow$S  transition has occurred, in synchronized flow that has emerged   at the bottleneck  
   speed peaks appear  (speed peaks 1 and 2 in Fig.~\ref{FSF_KKSW_traj3_F_A} (b, d)) (see item (iii) of 
   the common microscopic features of traffic breakdown  of Sec.~\ref{Mic_FS_S}). The physics of the speed peaks is the same as that discussed in
   Secs.~\ref{Synch_KKSW_S} and~\ref{Peaks_KKSW_S}.  The speed peaks lead to the emergence of dissolving
  speed waves in the synchronized flow (Fig.~\ref{FSF_KKSW_traj3_F_A}); the dissolving waves have also qualitatively the same microscopic features
  as shown  in Fig.~\ref{SFS_dis_onramp2_I_A}.
  
  \subsubsection{Return S$\rightarrow$F  transition due to  S$\rightarrow$F instability}

 A crucial difference of the case under consideration (Fig.~\ref{FSF_KKSW_traj1_F_A}) with traffic breakdown shown in Fig.~\ref{FS_Over_onrampLoc_A}
  becomes clear when we consider Fig.~\ref{FSF_KKSW_traj4_F_A}. We find that synchronized flow exists for a few minutes
  only: A speed peak (speed peak 3 in Fig.~\ref{FSF_KKSW_traj4_F_A}) occurs at the downstream front of this synchronized flow
  that initiates an S$\rightarrow$F  instability at the bottleneck.
  The S$\rightarrow$F  instability interrupts the formation of a congested pattern
  at the bottleneck.
  
  Indeed, due to the S$\rightarrow$F instability, rather than an WSP occurs, as this is realized in Fig.~\ref{FS_Over_onrampLoc_A},
a localized region of synchronized flow {\it departs  from the bottleneck}: 
The downstream front and the upstream front of this   synchronized flow 
(labeled by $\lq\lq$downstream front" and $\lq\lq$upstream front" in Figs.~\ref{FSF_KKSW_traj1_F_A}    (c, d) and~\ref{FSF_KKSW_traj7_F_A} (a)) propagate upstream from the bottleneck.
While propagating upstream from the bottleneck,   synchronized flow
dissolves over time.
Due to the occurrence of such a dissolving synchronized flow, the minimum speed $v_{\rm min}(t)$  within the permanent disturbance 
firstly decreases and then increases over time (trajectories 32--41 in Fig.~\ref{FSF_KKSW_traj1_F_A}    (e)).

  The physics of the S$\rightarrow$F instability is the same as disclosed in Sec.~\ref{Wave_D_KKSW_S}.
In particular, the S$\rightarrow$F  instability  leads to a growing wave of increase in speed within synchronized flow
  (labeled by $\lq\lq$growing speed wave" in 
  Fig.~\ref{FSF_KKSW_traj4_F_A}). The growth of the speed wave is realized due to 
  over-acceleration effect (Fig.~\ref{FSF_KKSW_traj7_F_A}) whose physics is the same as that discussed in Sec.~\ref{Wave_D_KKSW_S}.

   \subsection{Spatiotemporal $\lq\lq$overlapping" speed adaptation and over-acceleration effects  
  \label{Mic_Perm_Dis_S}}

          \begin{figure} 
 \begin{center}
\includegraphics*[width=8 cm]{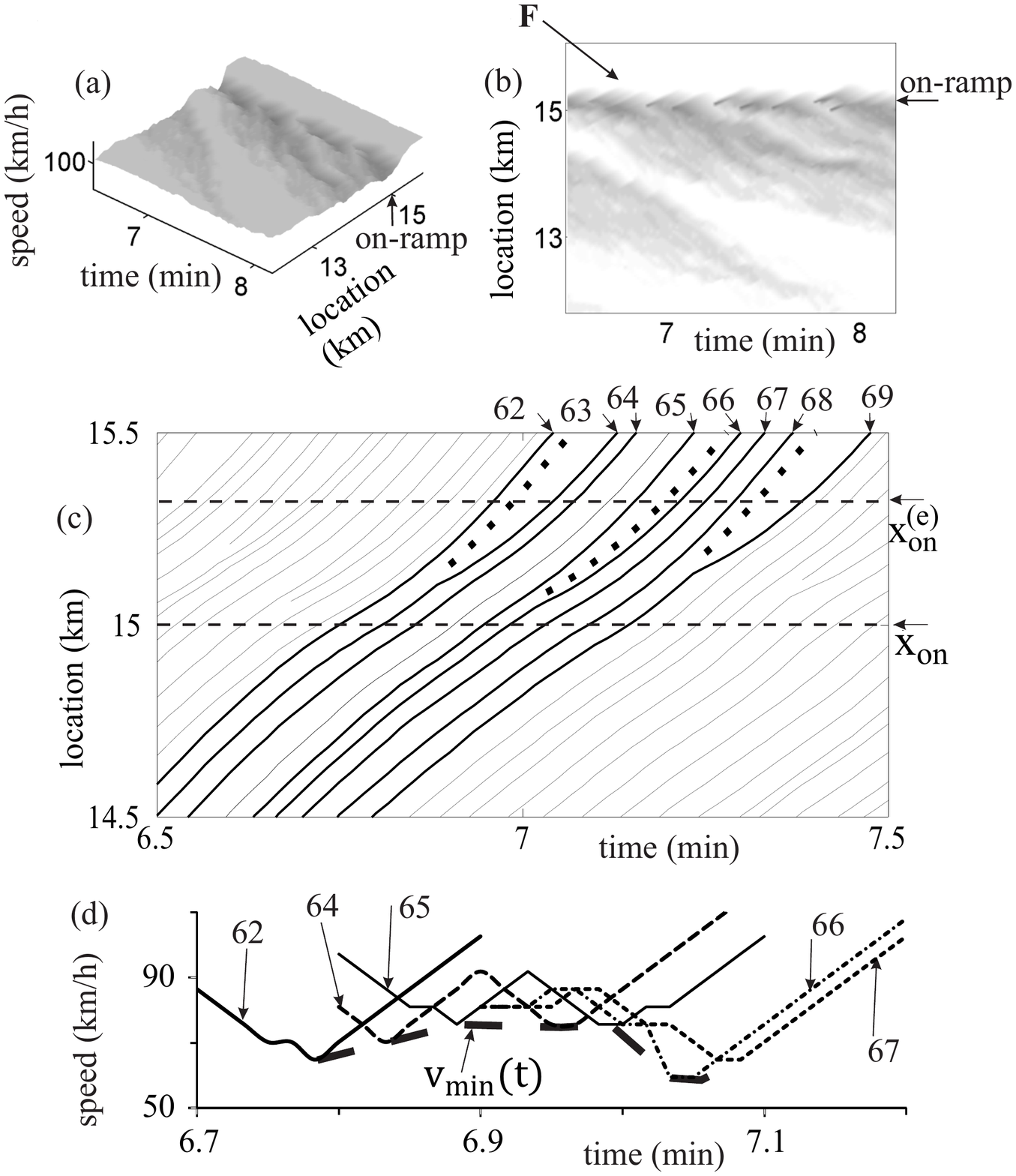}
 \end{center}
\caption{Simulations of dynamics of
 permanent disturbance at on-ramp bottleneck: 
(a, b) Speed in space and time (a) and the same speed data
presented  by regions with variable shades of gray (in white regions the speed
is  equal to or higher than 100  km/h, in black regions the speed is equal to 20 km/h) (b) for  a short time interval related to $t<T^{\rm (B)}_{1}$ in Fig.~\ref{FS_delay_onramp_A} (a). 
(c) Fragment of vehicle trajectories in space and time. (d) Microscopic vehicle speeds along trajectories as time functions
labeled by the same numbers as those in (c).
}
\label{FF_KKSW_traj1_F_A}
\end{figure}

  \begin{figure} 
\begin{center}
\includegraphics*[width=7.5 cm]{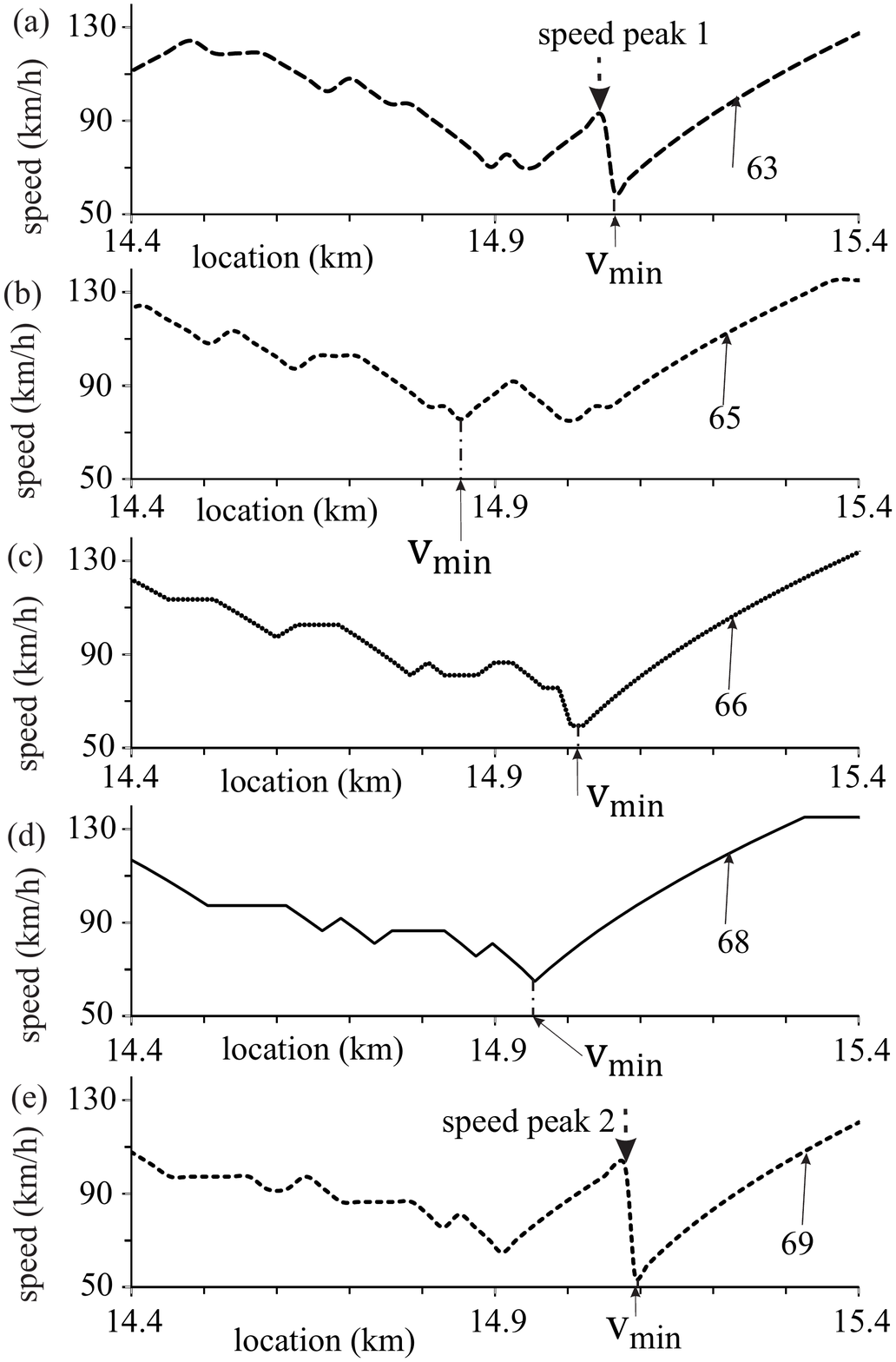}
 \end{center}
\caption{Simulations of the occurrence of   speed peaks and their evolution   within a permanent disturbance at on-ramp bottleneck:  
 (a--e) Microscopic vehicle speeds along trajectories as  
road location-functions  
labeled by the same numbers as those in Fig.~\ref{FF_KKSW_traj1_F_A} (c).
}
\label{FF_KKSW_traj3_1_F_A}
\end{figure} 

    \begin{figure} 
\begin{center}
\includegraphics*[width=7.5 cm]{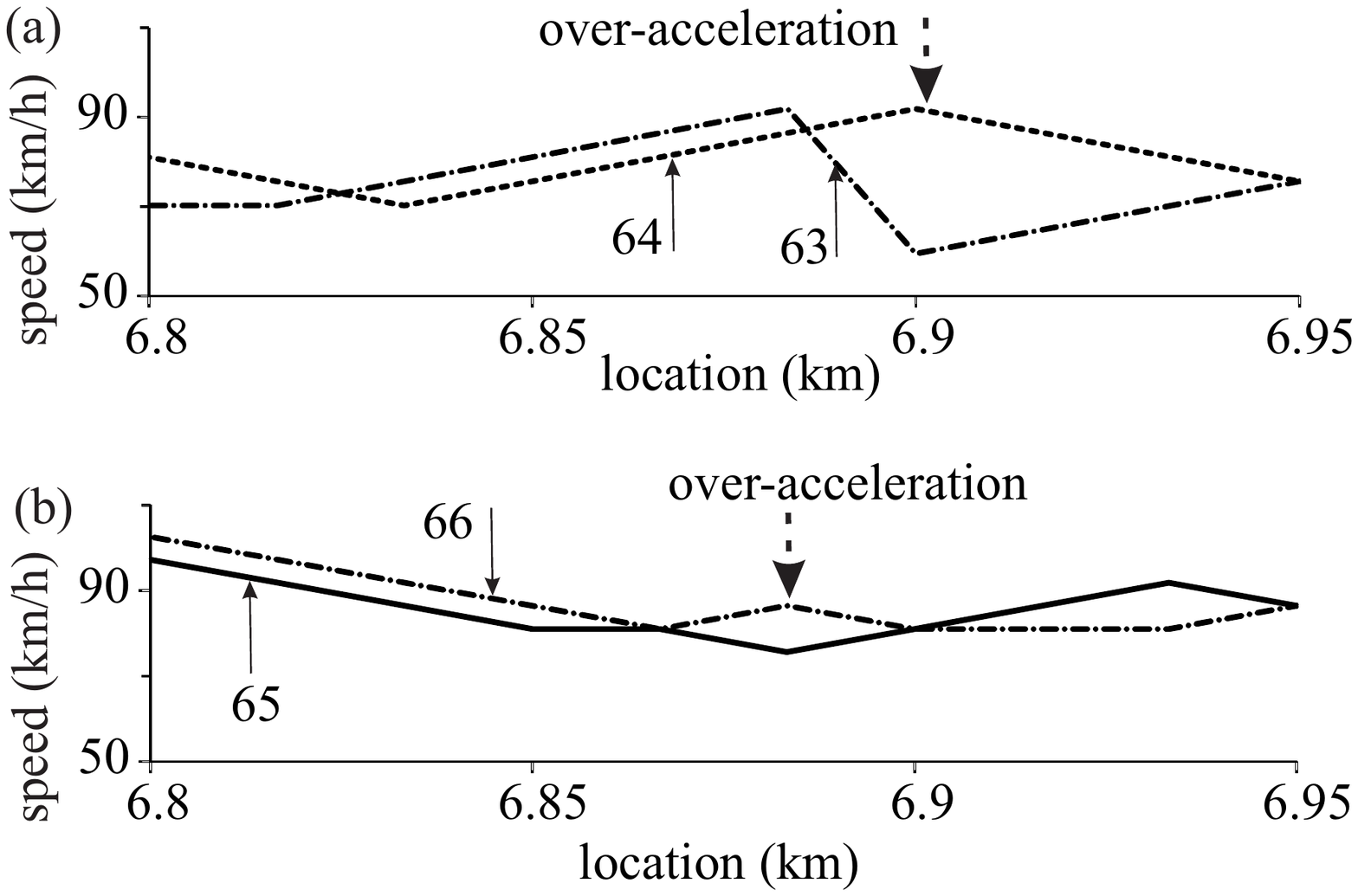}
 \end{center}
\caption{Simulations of over-acceleration   within a permanent disturbance at on-ramp bottleneck:  
 (a, b) Microscopic vehicle speeds along trajectories as  
road location-functions  
labeled by the same numbers as those in Fig.~\ref{FF_KKSW_traj1_F_A} (c).
}
\label{FF_KKSW_traj5_F_A}
\end{figure}

 During time delay $0< t <  T^{\rm (B)}_{1}$  of the breakdown
 (Fig.~\ref{FS_delay_onramp_A} (a)), there are also time intervals within which there is no noticeable time lag
   between the beginning of the F$\rightarrow$S transition    
   and  the S$\rightarrow$F instability due to over-acceleration. In this case, rather than to distinguish 
    a sequence of F$\rightarrow$S$\rightarrow$F transitions within the permanent speed disturbance at the bottleneck, we find  
 a  spatiotemporal $\lq\lq$overlapping" of the speed adaptation and over-acceleration effects.  
 
  In this case (Figs.~\ref{FF_KKSW_traj1_F_A}--\ref{FF_KKSW_traj5_F_A}), there is an upstream front of the permanent disturbance within which vehicles on the main road decelerate
   to a smaller   speed due to slower moving vehicles that merge from the on-ramp. Vehicles upstream of the upstream front of the disturbance move
   at their maximum free flow speed $v_{\rm free}$.
   There is also a downstream front of the disturbance within which
   vehicles accelerate to the maximum free flow speed $v_{\rm free}$ (Fig.~\ref{FF_KKSW_traj1_F_A}).
 We have found that the  distribution of the speed within the permanent disturbance
   exhibits a   complex spatiotemporal dynamics: 
   
   (i) The value of the minimum   speed $v_{\rm min}$ within the  disturbance changes randomly
   over time (Fig.~\ref{FF_KKSW_traj1_F_A} (d)). 
   
   (ii) This speed minimum occurs randomly at different   road locations (Fig.~\ref{FF_KKSW_traj3_1_F_A}). 
   
   (iii)
   There can be several speed maxima within the disturbance whose locations are also change randomly (Fig.~\ref{FF_KKSW_traj3_1_F_A}).

   This complex dynamics of the permanent speed disturbance at the bottleneck is explained as follows. 
   As in the fully developed synchronized flow (Fig.~\ref{SFS_dis_onramp_I_A} (a)), within the permanent speed disturbance there is a sequence of speed peaks that occur randomly
   at the downstream front of the permanent speed disturbance 
   (labeled by $\lq\lq$speed peak 1" and $\lq\lq$speed peak 2" in Fig.~\ref{FF_KKSW_traj3_1_F_A} (a, c)). The physics of these speed peaks
   is the same as that already explained in Sec.~\ref{Synch_KKSW_S}.
   
   Due to the speed peaks, regions of  increase in speed appears propagating upstream within the disturbance.
   Within the regions of speed increase, the over-acceleration effect occurs that
   prevents the upstream propagation of the upstream front of synchronized flow due to the speed adaptation.
   Examples of  the over-acceleration effect are shown in Fig.~\ref{FF_KKSW_traj5_F_A}.
   Vehicle 63 accelerates firstly and then begins to decelerates strongly  
   (Fig.~\ref{FF_KKSW_traj5_F_A} (a); see also speed peak 1 shown  in Fig.~\ref{FF_KKSW_traj3_1_F_A} (a)). However, the following vehicle 64 continues to accelerate even when
   preceding vehicle 63 decelerates strongly (labeled by $\lq\lq$over-aceleration" in (Fig.~\ref{FF_KKSW_traj5_F_A} (a))).  
    In another example, the following vehicle 66 begins to accelerate
   when the preceding vehicle 65 starts to decelerate (labeled by $\lq\lq$over-aceleration" in (Fig.~\ref{FF_KKSW_traj5_F_A} (b))). 
   
   These over-acceleration effects can be considered
   short time S$\rightarrow$F instabilities that
   increase the speed within the permanent speed disturbance. These short time S$\rightarrow$F instabilities prevent 
   a continuous propagation of the upstream front of the permanent speed disturbance, i.e., they prevent traffic breakdown at the bottleneck.  
   Therefore, rather than traffic breakdown (Fig.~\ref{FS_Over_onrampLoc_A} (a, d)) resulting in the formation of ${\rm WSP_{1}}$ (Fig.~\ref{FS_delay_onramp_A} (a)),
   the permanent speed disturbance persists at the bottleneck (Fig.~\ref{FF_KKSW_traj1_F_A} (a, d)). Thus, the   competition between speed adaptation and over-acceleration  
   determines a random time delay of traffic breakdown at the bottleneck independent on whether
     sequences of F$\rightarrow$S$\rightarrow$F transitions (Sec.~\ref{Mic_FSF_S}) can be distinguished or not within the permanent speed disturbance at the bottleneck.

       \section{General character of   effect of S$\rightarrow$F instability on   nucleation nature of traffic breakdown \label{Gen_S}}
       
       In Sec.~\ref{Nuc_Sec}, we have found that an S$\rightarrow$F instability is the origin of   
       sequences of F$\rightarrow$S$\rightarrow$F transitions at the bottleneck. In its turn, the F$\rightarrow$S$\rightarrow$F transitions
         is the reason of the nucleation nature of traffic breakdown. In other words, the S$\rightarrow$F instability governs the nucleation character of traffic breakdown at the bottleneck.

          \begin{figure} 
\begin{center}
\includegraphics*[width=7.5 cm]{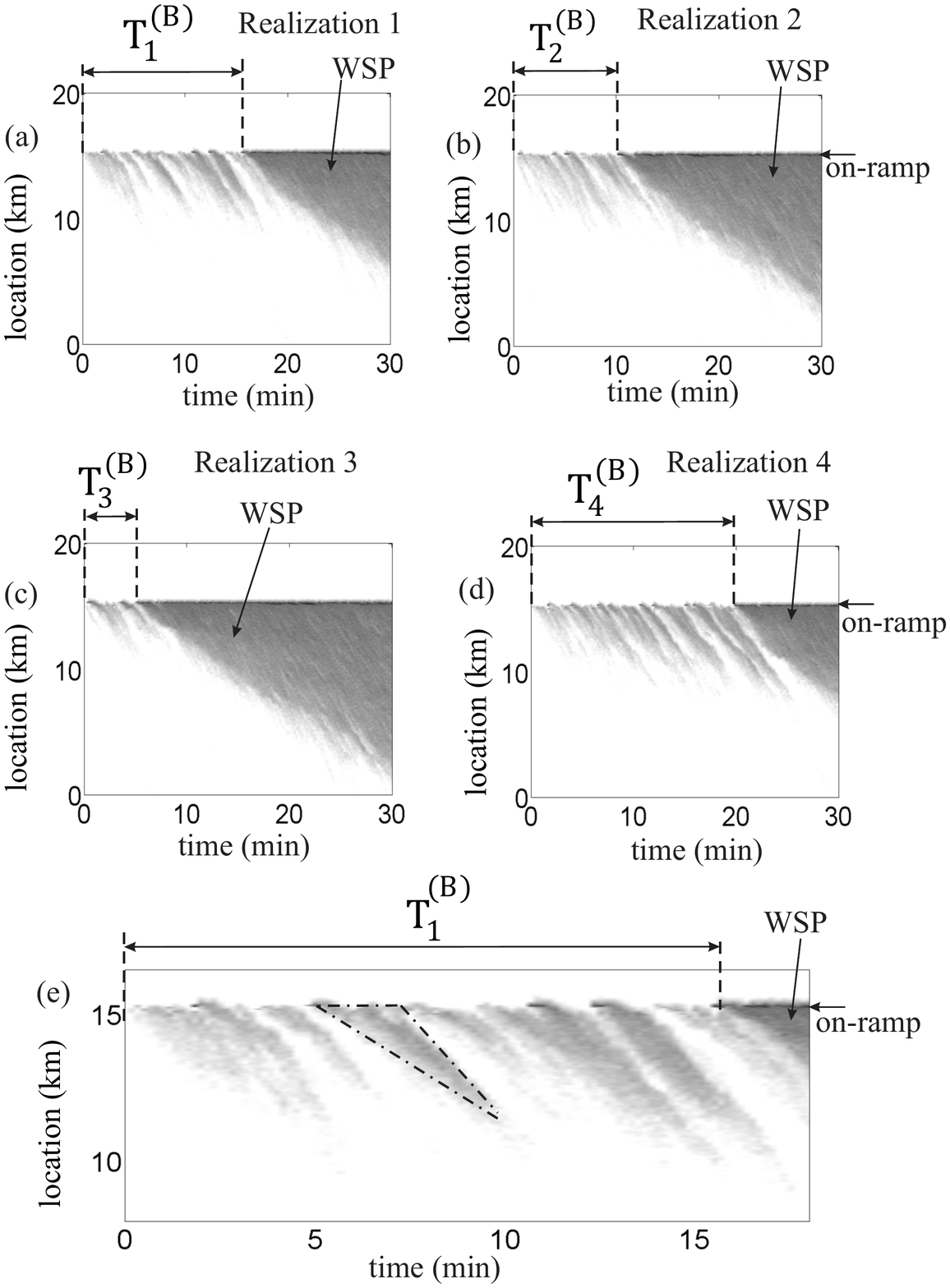}
 \end{center}
\caption{Random time delay of traffic breakdown (F$\rightarrow$S transition) at on-ramp bottleneck at a larger on-ramp inflow rate $q_{\rm on}=$ 480   vehicles/h than that in Fig.~\ref{FS_delay_onramp_A}:
(a--d)   Speed in space and time  for four different simulation realizations (runs)
presented  by regions with variable shades of gray (in white regions the speed
is  equal to or higher than 110  km/h, in black regions the speed is zero). (e) Speed in space and time  
presented  by regions with variable shades of gray   illustrating F$\rightarrow$S$\rightarrow$F transitions during
 time delay of traffic breakdown related to realization 1  in   (a). 
Time delays of traffic breakdown $T^{\rm (B)}$ in different simulation realizations 1--4 are
$T^{\rm (B)}_{1}=$ 16 min (a), $T^{\rm (B)}_{2}=$ 11 min (b),  $T^{\rm (B)}_{3}=$ 6 min (c), and $T^{\rm (B)}_{4}=$ 20 min (d).  The flow rate in free flow
upstream of the bottleneck is the same as that in Fig.~\ref{FS_delay_onramp_A}:
$q_{\rm in}=$  1406   vehicles/h.
}
\label{FS_delay_onramp_480_A}
\end{figure}

 However, when the on-ramp inflow rate $q_{\rm on}$ increases considerably, {\it no} S$\rightarrow$F instability is observed
         within   congested patterns (WSPs) that emerge after traffic breakdown has occurred at the bottleneck   (Fig.~\ref{FS_delay_onramp_480_A} (a--d))~\cite{General}. 
         This is in contrast with the WSPs shown in Fig.~\ref{FS_delay_onramp_A}.
         
 Due to the increase in $q_{\rm on}$, the mean speed of synchronized flow in  WSPs shown in Fig.~\ref{FS_delay_onramp_480_A} (a--d)
that emerge at the bottleneck after traffic breakdown has occurred
becomes smaller than the mean speed of synchronized flow in  WSPs shown in Fig.~\ref{FS_delay_onramp_A} (a, c).
We have found that also in the case of the WSPs shown in Fig.~\ref{FS_delay_onramp_480_A} (a--d)
 there are many random speed peaks  at the downstream front of synchronized flow; the speed peaks (not shown)     
  are qualitatively the same as those in Fig.~\ref{SFS_dis_onramp_I_A}.
 However, due to a smaller
  mean speed of synchronized flow in the WSPs, no S$\rightarrow$F instability can be initiated
  by these speed peaks during the whole  time of the observation of traffic flow $T_{\rm ob}=$ 30 min in    Fig.~\ref{FS_delay_onramp_480_A}:
   The speed peaks initiate  only dissolving speed waves in synchronized flow (not shown) that are qualitatively similar to those shown in
   Figs.~\ref{SFS_dis_onramp_I_A} (b) and~\ref{SFS_dis_onramp2_I_A} found for a smaller on-ramp inflow rate.
   
   Although there are no S$\rightarrow$F instabilities within the WSPs,
   we have found random time delays of traffic breakdown at the bottleneck (Fig.~\ref{FS_delay_onramp_480_A} (a--d)) that exhibit
   the same features as those in Fig.~\ref{FS_delay_onramp_A} (a--d).
  We have also found that there are sequences of F$\rightarrow$S$\rightarrow$F transitions  that are
         the reason for the existence of a random time delay of traffic breakdown. Each of the sequences of  F$\rightarrow$S$\rightarrow$F transitions
         (one of them is marked by dashed-dotted curves in 
         Fig.~\ref{FS_delay_onramp_480_A} (e)) exhibits qualitatively the same physical features as those found out in Sec.~\ref{Mic_FSF_S}.   
         
         In other words, the result of this article that
         the S$\rightarrow$F instability governs  the metastability of free flow  with respect to
traffic breakdown      at the bottleneck  exhibits a general character.
         The physics of this general result is as follows.

        (i) There are sequences of  F$\rightarrow$S$\rightarrow$F transitions at the bottleneck (Sec.~\ref{Mic_FSF_S}).
        On average, the F$\rightarrow$S$\rightarrow$F transitions cause  
           a permanent speed disturbance, i.e., a permanent decrease in speed in free flow 
          localized  at the bottleneck. The   permanent speed disturbance exhibits a complex dynamic behavior in space and time.

(ii) When a  decrease  in speed within the permanent speed disturbance in free flow
becomes randomly equal to or larger than some critical  decrease  in speed, the resulting F$\rightarrow$S  transition, i.e., the  upstream propagation of 
 the upstream front of the synchronized flow cannot be suppressed by the
S$\rightarrow$F instability. In this case as considered in Sec.~\ref{Mic_FS_S}, rather than a sequence of F$\rightarrow$S$\rightarrow$F transitions,
   a congested pattern emerges at the bottleneck (WSPs in Figs.~\ref{FS_delay_onramp_A} and~\ref{FS_delay_onramp_480_A}). Otherwise, when the local decrease  in speed
in free flow at the bottleneck is smaller than the critical one, the S$\rightarrow$F instability interrupts the development of the F$\rightarrow$S  transition: Rather than the congested pattern,
a sequence of the F$\rightarrow$S$\rightarrow$F transitions
occurs at the bottleneck.

(iii) There can be  a    time interval during which any  decrease in speed
within the permanent speed disturbance in free flow at the bottleneck is smaller than the critical one.
In this case, the S$\rightarrow$F instability   interrupts the development of each of the F$\rightarrow$S  transitions.  
 This   time interval is the time delay  $T^{\rm (B)}$
 of traffic breakdown (Figs.~\ref{FS_delay_onramp_A} and~\ref{FS_delay_onramp_480_A}).
  
  (iv) The time delay of traffic breakdown (Figs.~\ref{FS_delay_onramp_A} and~\ref{FS_delay_onramp_480_A} (a--d)) is a 
random value because the S$\rightarrow$F instability exhibits the nucleation nature:
 The S$\rightarrow$F instability occurs only if a large enough initial  increase in speed, which
  is equal to or larger than a critical increase in speed,  appears randomly within the emergent synchronized flow
 at the bottleneck.

 (v) The  critical increase in speed in synchronized flow, at which an S$\rightarrow$F instability occurs,
  depends on the     critical  decrease  in speed within the permanent speed disturbance in free flow at the bottleneck, at which traffic breakdown occurs: When  the S$\rightarrow$F instability 
  cannot interrupt the development of the F$\rightarrow$S  transition, a congested pattern is formed at the bottleneck.

 If the on-ramp inflow rate $q_{\rm on}$ increases,
  while the flow rate on the main road upstream of the bottleneck  $q_{\rm in}$ remains, we have found the following   effects: 
  
  1. Within synchronized flow of a   congested pattern at the bottleneck,
  the probability of the occurrence of the S$\rightarrow$F instability  decreases.
   Indeed,
   in contrast with the cases shown in Figs.~\ref{FS_delay_onramp_A} (a--d), there is no S$\rightarrow$F instability
   within WSPs in Figs.~\ref{FS_delay_onramp_480_A} (a--d). 
   
   2. The   mean time delay of traffic breakdown becomes shorter:
    The mean value of the time delay of traffic breakdowns shown in Fig.~\ref{FS_delay_onramp_480_A}
    is shorter than that in Fig.~\ref{FS_delay_onramp_A}.

 \section{Discussion  \label{Dis_S}}

 \subsection{Classical traffic flow instability versus S$\rightarrow$F instability of three-phase theory \label{GM_S}}
 
\begin{figure} 
\begin{center}
\includegraphics*[width=7.5 cm]{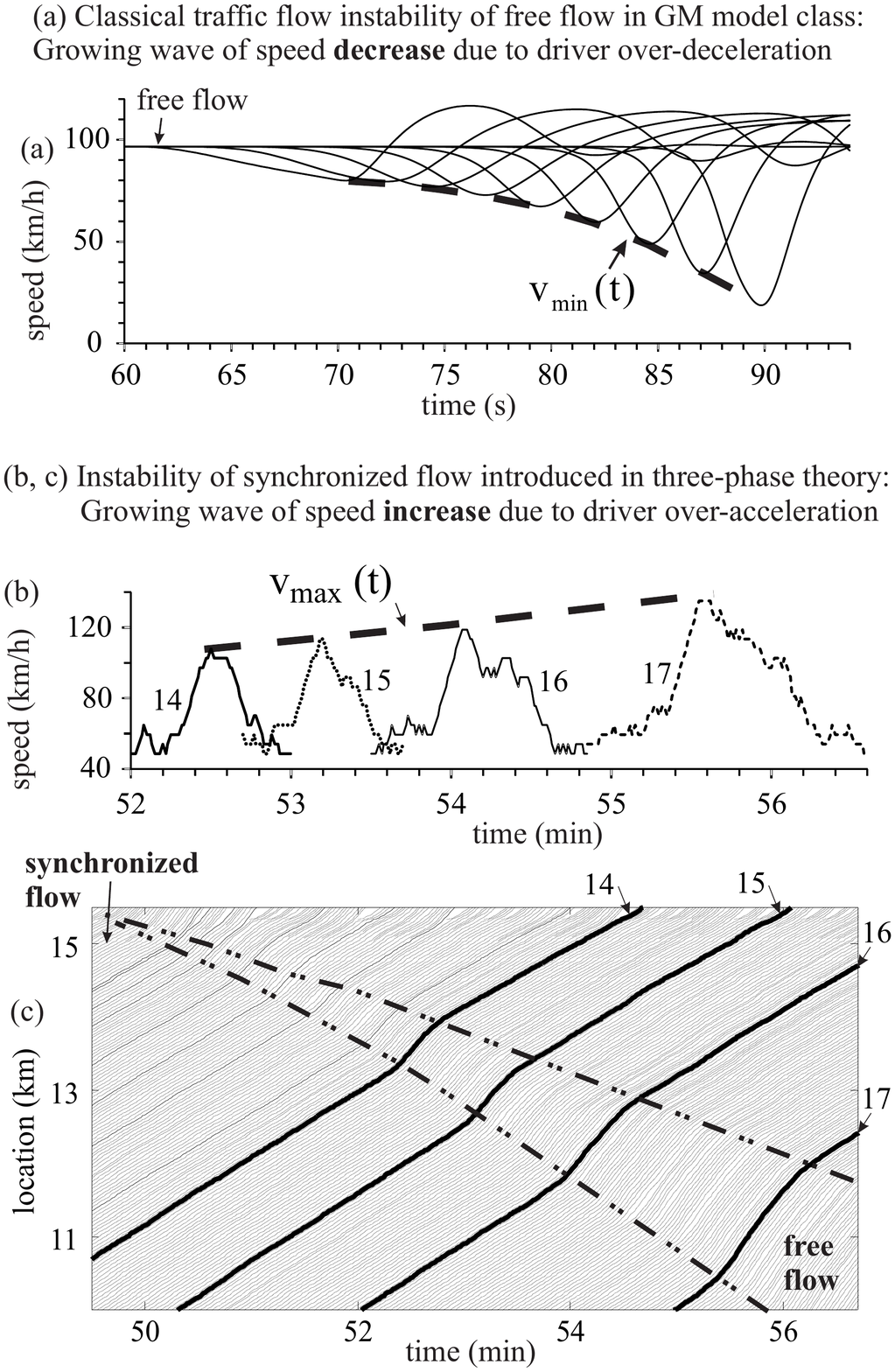}
 \end{center}
\caption{Classical traffic flow instability (a)~\cite{GH195910,Gazis1961A10,GH10,Chandler,KS,KS1959A,Newell1961,Newell1963A,Newell1981,Newell_Stoch,Gipps,Gipps1986,Wiedemann,Whitham1990,ach_Pay197110,ach_Pay197910,Stoc,Bando1995,ach_Kra10,fail_Nagatani1998A,fail_Nagatani1999B,ach_Helbing200010,ach_Aw200010,ach_Jiang2001A,Reviews,Reviews2} versus S$\rightarrow$F instability of three-phase theory (b, c):
(a) Vehicle trajectories as time-functions showing the well-known growing wave of {\it speed reduction} caused by classical traffic flow 
instability
 with simulations of optimal velocity model by Bando {\it et al.}~\cite{Bando1995,Bando_Model}.
(b, c) Vehicle trajectories as time-functions (b) (taken from Fig.~\ref{SF_Over_onramp2_3_I_A}) and in space and time
 showing the   growing wave of {\it speed increase} caused by S$\rightarrow$F instability (taken from Fig.~\ref{SFS_Over_onramp_I_A}).
}
\label{Instability_Com}
\end{figure}

 The basic difference between the classical traffic flow 
instability~\cite{GH195910,Gazis1961A10,GH10,Chandler,KS,KS1959A,Newell1961,Newell1963A,Newell1981,Newell_Stoch,Gipps,Gipps1986,Wiedemann,Whitham1990,ach_Pay197110,ach_Pay197910,Stoc,Bando1995,ach_Kra10,fail_Nagatani1998A,fail_Nagatani1999B,ach_Helbing200010,ach_Aw200010,ach_Jiang2001A,Reviews,Reviews2}
  and an S$\rightarrow$F instability of three-phase theory
  is as follows:
 The classical traffic instability is a growing wave of local {\it decrease} in speed in free flow
  (Fig.~\ref{Instability_Com} (a))~\cite{GH195910,Gazis1961A10,GH10,Chandler,KS,KS1959A,Newell1961,Newell1963A,Newell1981,Newell_Stoch,Gipps,Gipps1986,Wiedemann,Whitham1990,ach_Pay197110,ach_Pay197910,Stoc,Bando1995,ach_Kra10,fail_Nagatani1998A,fail_Nagatani1999B,ach_Helbing200010,ach_Aw200010,ach_Jiang2001A,Reviews,Reviews2}. 
  Contrary,
 the S$\rightarrow$F instability is  a growing wave of local {\it increase} in speed in synchronized  flow (Fig.~\ref{Instability_Com} (b, c)). 
 
The classical traffic flow 
instability~\cite{GH195910,Gazis1961A10,GH10,Chandler,KS,KS1959A,Newell1961,Newell1963A,Newell1981,Newell_Stoch,Gipps,Gipps1986,Wiedemann,Whitham1990,ach_Pay197110,ach_Pay197910,Stoc,Bando1995,ach_Kra10,fail_Nagatani1998A,fail_Nagatani1999B,ach_Helbing200010,ach_Aw200010,ach_Jiang2001A,Reviews,Reviews2}  should explain traffic breakdown through the driver reaction time
(time delay in driver over-deceleration).  However,   this classical traffic flow  instability leads to a phase transition from free flow to a wide moving jam (F$\rightarrow$J transition)~\cite{KK1993,Reviews2,Kerner_Review,KernerBook,KernerBook2}.
The classical instability has been incorporated in a huge number of traffic flow models~\cite{Reviews2,Kerner_Review}.
Contrary to the classical traffic flow instability, in real field traffic data, traffic breakdown is an F$\rightarrow$S transition.
A more detailed explanation why   the classical traffic flow instability have failed to explain real   traffic breakdown
can be found in~\cite{Kerner_Review}.

However, it should be noted that the classical traffic 
instability~\cite{GH195910,Gazis1961A10,GH10,Chandler,KS,KS1959A,Newell1961,Newell1963A,Newell1981,Newell_Stoch,Gipps,Gipps1986,Wiedemann,Whitham1990,ach_Pay197110,ach_Pay197910,Stoc,Bando1995,ach_Kra10,fail_Nagatani1998A,fail_Nagatani1999B,ach_Helbing200010,ach_Aw200010,ach_Jiang2001A,Reviews,Reviews2}
has also been used in three-phase theory  to  explain   a growing wave of local {\it decrease} in speed within synchronized flow
leading to the emergence of a wide moving jam(s) in synchronized flow (S$\rightarrow$J 
transition) (Fig.~\ref{TwoIn_Emp} (d))~\cite{Kerner1998B,KernerBook}.
Thus in three-phase theory, the emergence of wide moving jams is realized through a sequence of F$\rightarrow$S$\rightarrow$J transitions~\cite{Kerner1998B,KernerBook}.
   
    \subsection{Traffic breakdown without over-acceleration \label{No_Over_KKSW_S}}
  
  \begin{figure} 
\begin{center}
\includegraphics*[width=8 cm]{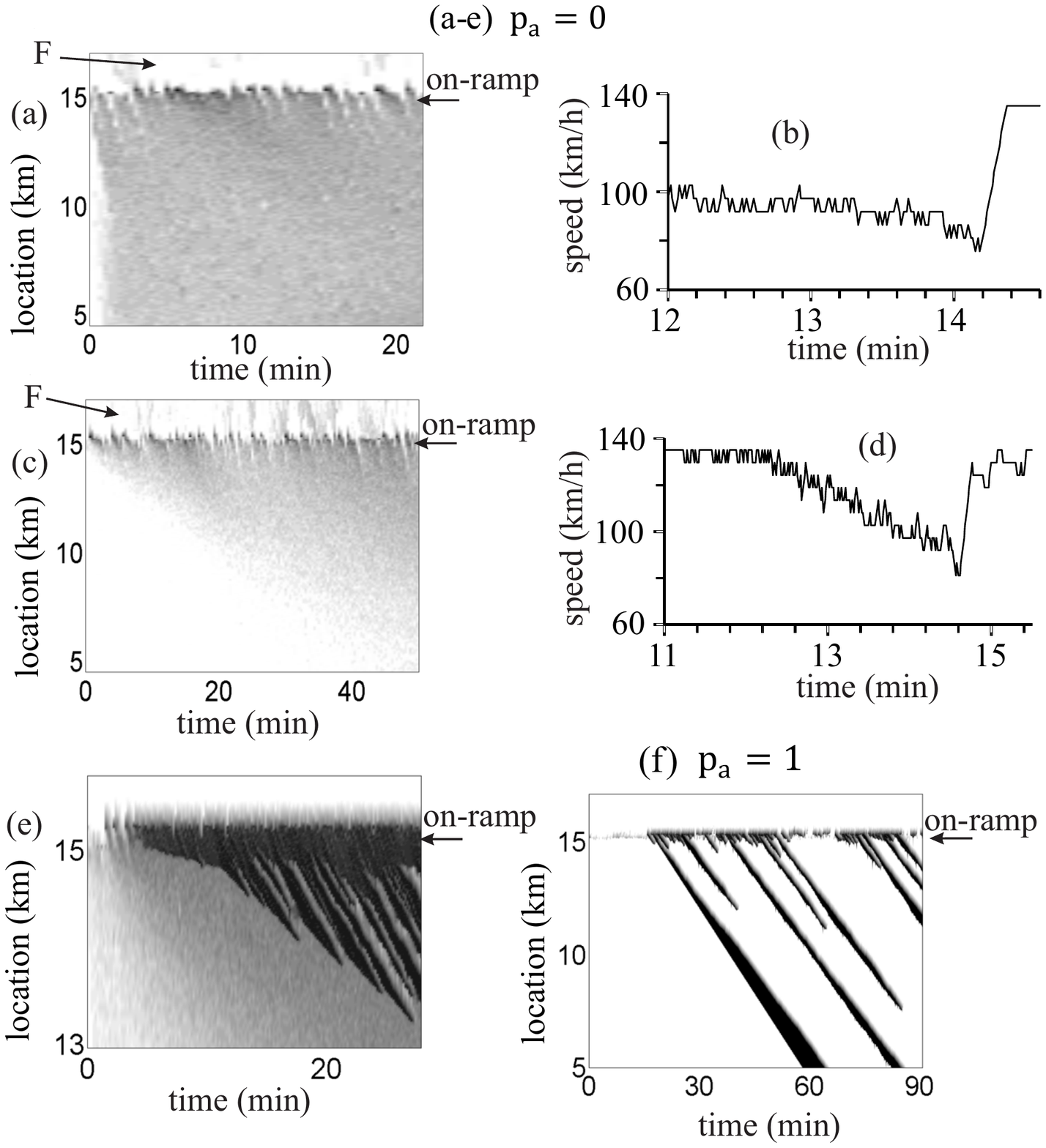}
 \end{center}
\caption{Simulations of traffic breakdown at on-ramp bottleneck with the KKSW CA model in which either  the
 probability of over-acceleration $p_{\rm a}=0$ (a--e) (Sec.~\ref{No_Over_KKSW_S}) or the
 probability of over-acceleration $p_{\rm a}=1$ (f) (Sec.~\ref{NoDelay_Over_KKSW_S}):  
(a, c, e, f) Speed in space and time  
presented  by regions with variable shades of gray (in white regions the speed
is  equal to or higher than 130  km/h (a, c) or 120 km/h (e, f), in black regions the speed is equal to 30 km/h (a, c) or  zero (e, f)).
(b, d)   Microscopic vehicle speeds along one of the vehicle trajectories moving within
  patterns in (a, c), respectively.  
  $q_{\rm on}=$ 360 (a--d, f) and 900 (e) vehicles/h,
$q_{\rm in}=$ 1800 (f),  1406     (a, b, e), and 1125 vehicles/h  (c, d). Other model parameters are the same as those given in caption to Fig.~\ref{SteadyStates}.
}
\label{pa_0_KKSW_A}
\end{figure}

When in (\ref{Overacceleration1_KKW})  the probability of over-acceleration $p_{\rm a}=0$, there is no over-acceleration in the KKSW CA model.
In this case,
  no S$\rightarrow$F instability is realized.
For this reason,  we find that  congested traffic emerges at the bottleneck without any delay. The
  downstream front of the pattern is fixed at the bottleneck (Fig.~\ref{pa_0_KKSW_A} (a, b)). When we decrease the flow rate on the main road,
  congested traffic occurs also without any time delay;
  due to smaller flow rate upstream,
  the upstream front of this congested traffic propagates slower only (Fig.~\ref{pa_0_KKSW_A} (c, d)).
  Because the downstream front of the congested traffic is fixed at the bottleneck we can call it as $\lq\lq$synchronized flow".

   In other words, features of the synchronized flow shown in Fig.~\ref{pa_0_KKSW_A} (a--d) contradict
   the nucleation nature of traffic breakdown (F$\rightarrow$S transition)
  found in real field traffic data. Thus over-acceleration is needed to simulate
  the nucleation nature of an F$\rightarrow$S transition
  of real   traffic.

  The absent of over-acceleration ($p_{\rm a}=0$) does not affect the slow-of-start rule used in the KKSW CA model.
  Therefore, we can expect that an S$\rightarrow$J  instability can occur within synchronized flow leading to the emergence of a wide moving jam(s). Indeed,
  when we increase the on-ramp inflow rate, so that the mean speed in synchronized flow decreases considerably,
  moving jams emerge in this dense synchronized flow (Fig.~\ref{pa_0_KKSW_A} (e)).

 \subsection{Traffic breakdown without time delay of over-acceleration \label{NoDelay_Over_KKSW_S}}

The necessity of the existence of a finite time delay in over-acceleration to simulate an S$\rightarrow$F instability and, therefore,
the nucleation features of traffic breakdown becomes more clear, if we assume that over-acceleration occurs with probability $p_{\rm a}=$1, i.e.,
without any time delay. 

Because such a limit case   is not attained with   the KKSW CA model (\ref{nonadaptation1})--(\ref{S_Gap}), we should make the following changes in the model:
When $p_{\rm a}=$1,  model step (c) (Eq. (\ref{Overacceleration1_KKW})) is satisfied with probability 1. In step (f) (Eq. (\ref{Randomization_KKSW})), rather than Eq. (\ref{rand_p}), 
the following formula is used
\begin{equation}
r<p.
\label{rand_p_lim}
\end{equation}

We have found that when over-acceleration occurs without time delay, such over-acceleration prevents speed adaptation within
2D-states of synchronized flow. Therefore, synchronized flow states are not realized. In other words,
there are no S$\rightarrow$F instability and no time-delayed F$\rightarrow$S transition  in this model. In general, 
such model exhibits qualitatively the same features of traffic breakdown at the bottleneck as those of the NaSch  CA model~\cite{Stoc2,Schadschneider_Book}:
Traffic breakdown is governed by the classical traffic flow instability of the GM model class (Sec.~\ref{GM_S}) leading to
  a well-known time-delayed F$\rightarrow$J transition (Fig.~\ref{pa_0_KKSW_A} (f)).

 \subsection{General   microscopic features of the S$\rightarrow$F instability  \label{KKSW_KKl_S}} 
 
 Microscopic features of the S$\rightarrow$F instability derived  above based on a study of
  the KKSW CA model exhibit general character, i.e., they are independent
 on   specific properties of   the KKSW CA model. To prove this statement, we show that
qualitatively the same features of the S$\rightarrow$F instability can be derived with simulations of the Kerner-Klenov stochastic three-phase model 
 of~\cite{KKl,KKl2003A,KKl2009A}. We use a discrete in space model version of~\cite{KKl2009A} for a single lane road with an on-ramp bottleneck
 (Appendix~\ref{App}).  
 
\subsubsection{Nucleation features of S$\rightarrow$F instability  \label{KKSW_KKl_N_S}} 
 
\begin{figure} 
\begin{center}
\includegraphics*[width=8 cm]{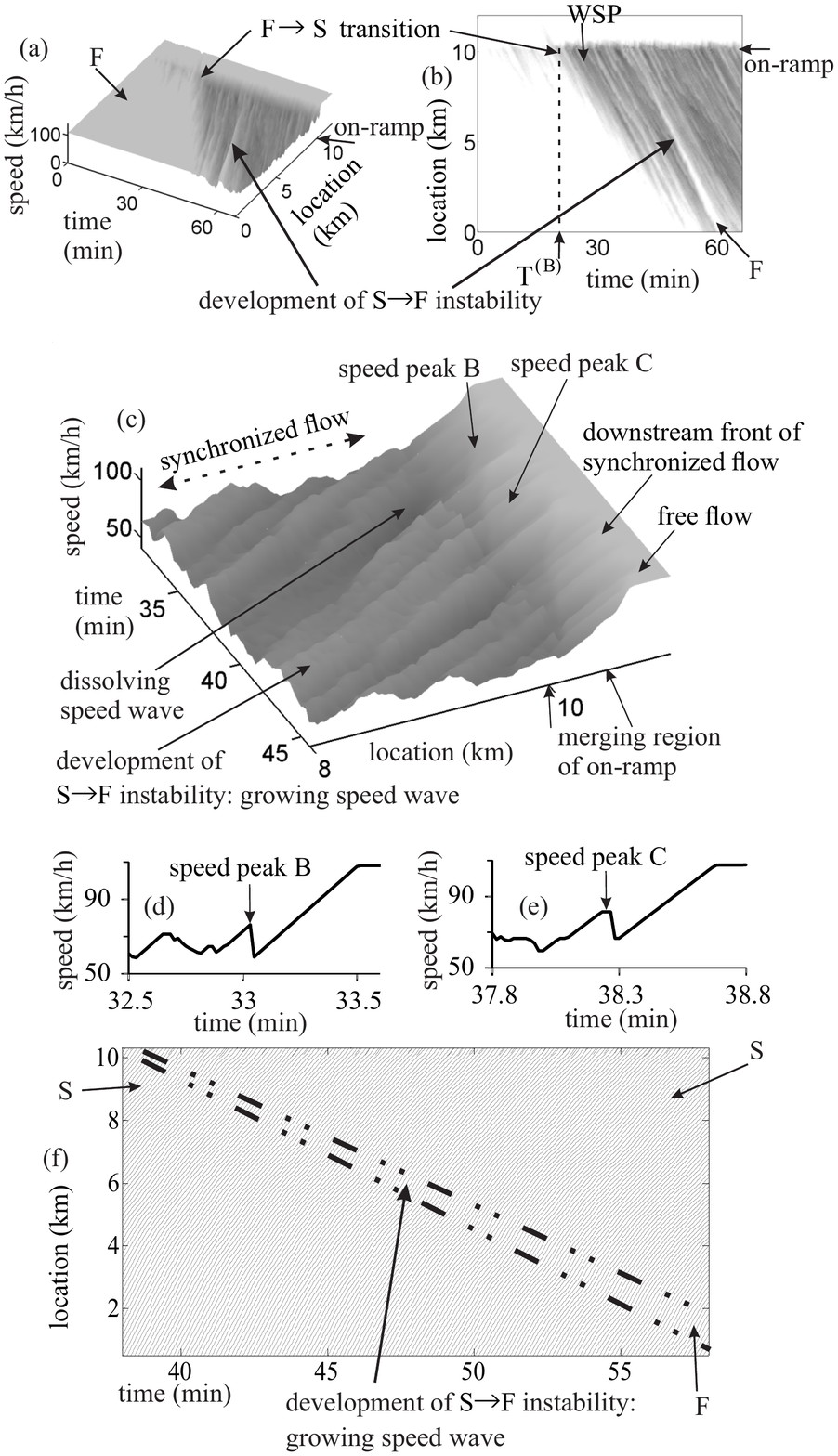}
 \end{center}
\caption{Simulations of
speed peaks at   downstream front of synchronized flow  
and   S$\rightarrow$F instability    at on-ramp bottleneck on single-lane road with the Kerner-Klenov model
(Tables~\ref{table_CA}--\ref{table1} of Appendix~\ref{App}):
(a, b) Speed in space and time (a) and the same   data
presented  by regions with variable shades of gray (in white regions the speed
is  equal to or higher than 100  km/h, in black regions the speed is zero) (b). 
(c) Speed in space and time for   time   $t>T^{\rm (B)}$  within synchronized flow of WSP; 
two of the speed peaks  in (c) are   marked by   $\lq\lq$speed peak B"
and  $\lq\lq$speed peak C".
(d, e) Microscopic (single-vehicle)   speeds 
along vehicle trajectories as   time-functions showing   speed peak B (d) leading to a dissolving
speed wave (labeled by $\lq\lq$dissolving
speed wave" in (b))
and speed peak C (e) initiating   a growing
speed wave (labeled by $\lq\lq$development of S$\rightarrow$F instability: growing
speed wave" in (b)).  The physics of  
speed peaks B and C is the same as that for
 speed peaks shown in Figs.~\ref{SFS_dis_onramp_I_A} (a, c) and~\ref{SF_Over_onramp2_p_I_A} (b):  
   vehicles shown in (d, e), which 
   begin to accelerate  at the downstream front of synchronized flow, 
have to interrupt their acceleration and to decelerate
due to   vehicles merging from the on-ramp onto the main road.     
 $x_{\rm on}=$ 10 km and $x^{\rm (e)}_{\rm on}=10.3$ km
are, respectively, the beginning and the end of the merging region of the on-ramp.
(f) Fragment of vehicle trajectories in space and time related to (a, b) (each 5th vehicle is shown);
 bold dashed-dotted curves in (f) mark the development of S$\rightarrow$F instability
in synchronized flow.
F -- free flow, S -- synchronized flow, WSP -- widening synchronized flow pattern.
$q_{\rm on}=$ 170   vehicles/h,
$q_{\rm in}=$  2278   vehicles/h. 
Other model parameters are given in Tables~\ref{table_parameters} and~\ref{table_parameters_bottlenecks}. 
 }
\label{SFS_dis_onramp_I_KKl}
\end{figure}

 (i)   As   in Fig.~\ref{SFS_Over_onramp_I_A} (a, b), after
 traffic breakdown (F$\rightarrow$S transition) has occurred at the bottleneck,
 synchronized flow emerges whose downstream front is localized at the bottleneck (Fig.~\ref{SFS_dis_onramp_I_KKl} (a, b)). A random
 sequence of speed peaks  
  appears at the downstream front of synchronized flow at the bottleneck
   (Fig.~\ref{SFS_dis_onramp_I_KKl} (c); compare with Fig.~\ref{SFS_dis_onramp_I_A} (a)). The speed peaks 
  are disturbances of   increase in speed
   in synchronized flow within which the microscopic (single-vehicle) speed is higher than the average
  synchronized flow speed (Fig.~\ref{SFS_dis_onramp_I_KKl} (d, e); compare with Figs.~\ref{SF_Over_onramp2_p_I_A} (b)
  and~\ref{SFS_dis_onramp_I_A} (c)).  
  
 (ii) As   in Fig.~\ref{SFS_dis_onramp_I_A} (b),
 small speed peaks (small disturbances of   increase in speed) in synchronized flow lead to   dissolving speed waves of 
 increase in speed in synchronized flow  ($\lq\lq$dissolving
  speed wave" in Fig.~\ref{SFS_dis_onramp_I_KKl} (c)). In this case, no
S$\rightarrow$F instability occurs. 

(iii)
Only when a   speed peak with a large enough increase in speed occurs randomly at the downstream front of synchronized flow at the bottleneck, the speed peak
   initiates the S$\rightarrow$F instability: A growing speed wave of increase in speed occurs in synchronized flow whose growth leads
   to an S$\rightarrow$F transition ($\lq\lq$growing speed wave" in Fig.~\ref{SFS_dis_onramp_I_KKl}  (c, f);
   compare with Fig.~\ref{SFS_Over_onramp_I_A} (e)).   As shown with simulations of the KKSW CA model
    in Fig.~\ref{SF_Over_onramp2_2_I_A}, simulations with the Kerner-Klenov model confirm (not shown) that
   the S$\rightarrow$F instability occurs due to the over-acceleration effect.
   
   The behavior of disturbances of   increase in speed in synchronized flow (items (ii) and (iii)) proves  
    the nucleation nature of the S$\rightarrow$F instability. 
 
 \subsubsection{S$\rightarrow$F instability as origin
  of nucleation nature of traffic breakdown at highway bottlenecks  \label{KKSW_KKl_O_S}}
  
\begin{figure} 
\begin{center}
\includegraphics*[width=8 cm]{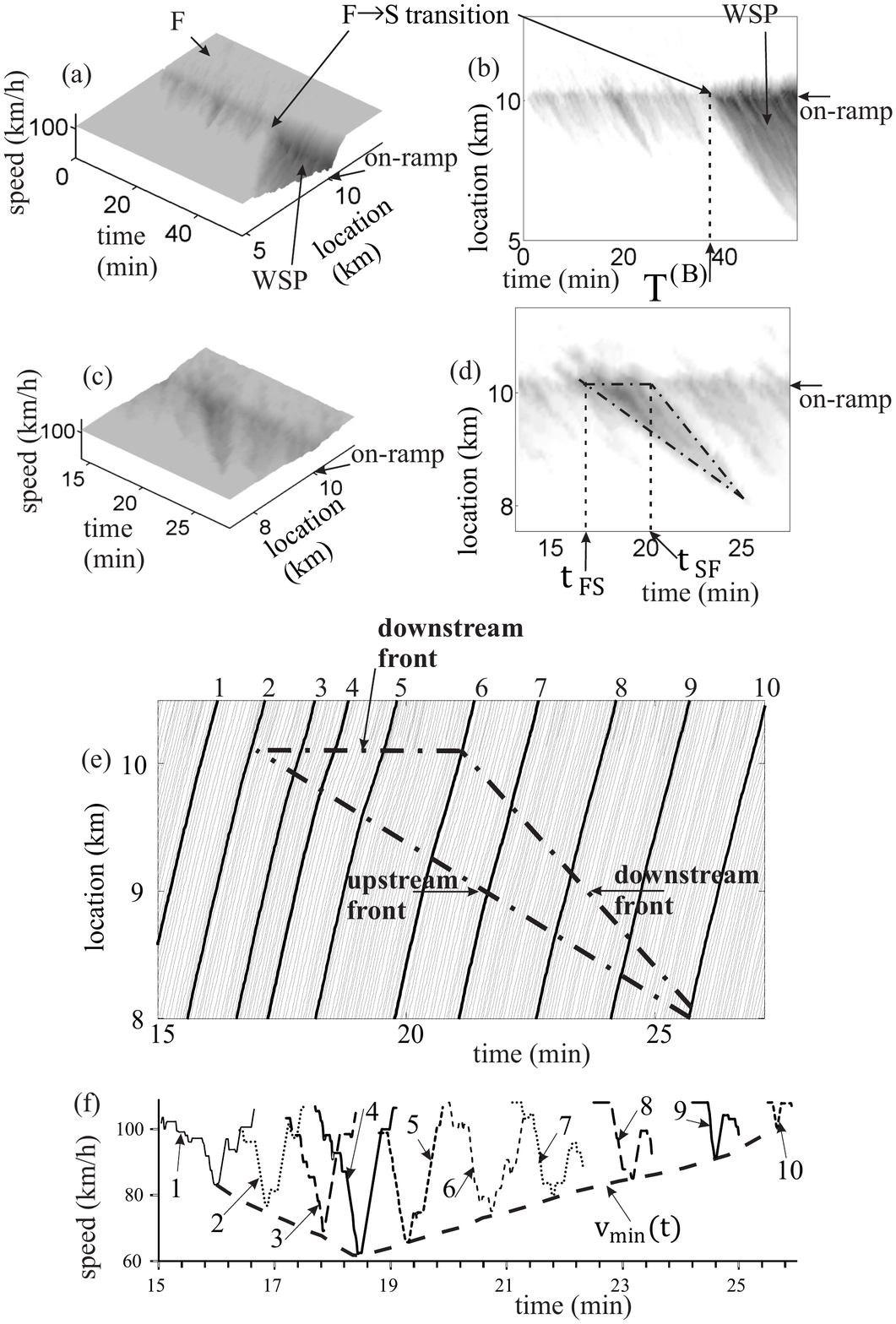}
 \end{center}
\caption{Simulations of F$\rightarrow$S$\rightarrow$F transitions within a permanent speed disturbance at on-ramp bottleneck on single-lane road with the Kerner-Klenov model
(Tables~\ref{table_CA}--\ref{table1} of Appendix~\ref{App}): 
(a, b) Speed in space and time  (a) and the same data
presented  by regions with variable shades of gray (in white regions the speed
is  equal to  or higher than 105  km/h, in black regions the speed is equal to 0 km/h).
(c, d) Speed in space and time (c) and the same  data
presented  by regions with variable shades of gray (d) (in white regions the speed
is  equal to  or higher than 100  km/h, in black regions the speed is equal to 20 km/h)  for  a short time interval in  (a, b). 
(e) Fragment of vehicle trajectories in space and time related to (c, d). (f) Microscopic vehicle speeds along trajectories as time functions
labeled by the same numbers as those in (e). In (d, e), dashed-dotted lines mark 
emergent synchronized flow that dissolves due to S$\rightarrow$F instability
(labels $\lq\lq$downstream front"
and $\lq\lq$upstream front" show boundaries of the synchronized flow region).
F -- free flow,   WSP -- widening synchronized flow pattern.
$q_{\rm on}=$ 320   vehicles/h,
$q_{\rm in}=$  2000   vehicles/h. 
Other model parameters are given in Tables~\ref{table_parameters} and~\ref{table_parameters_bottlenecks}. 
}
\label{FSF_KKl_traj1_F_A}
\end{figure}

\begin{figure} 
 \begin{center}
\includegraphics*[width=7.5 cm]{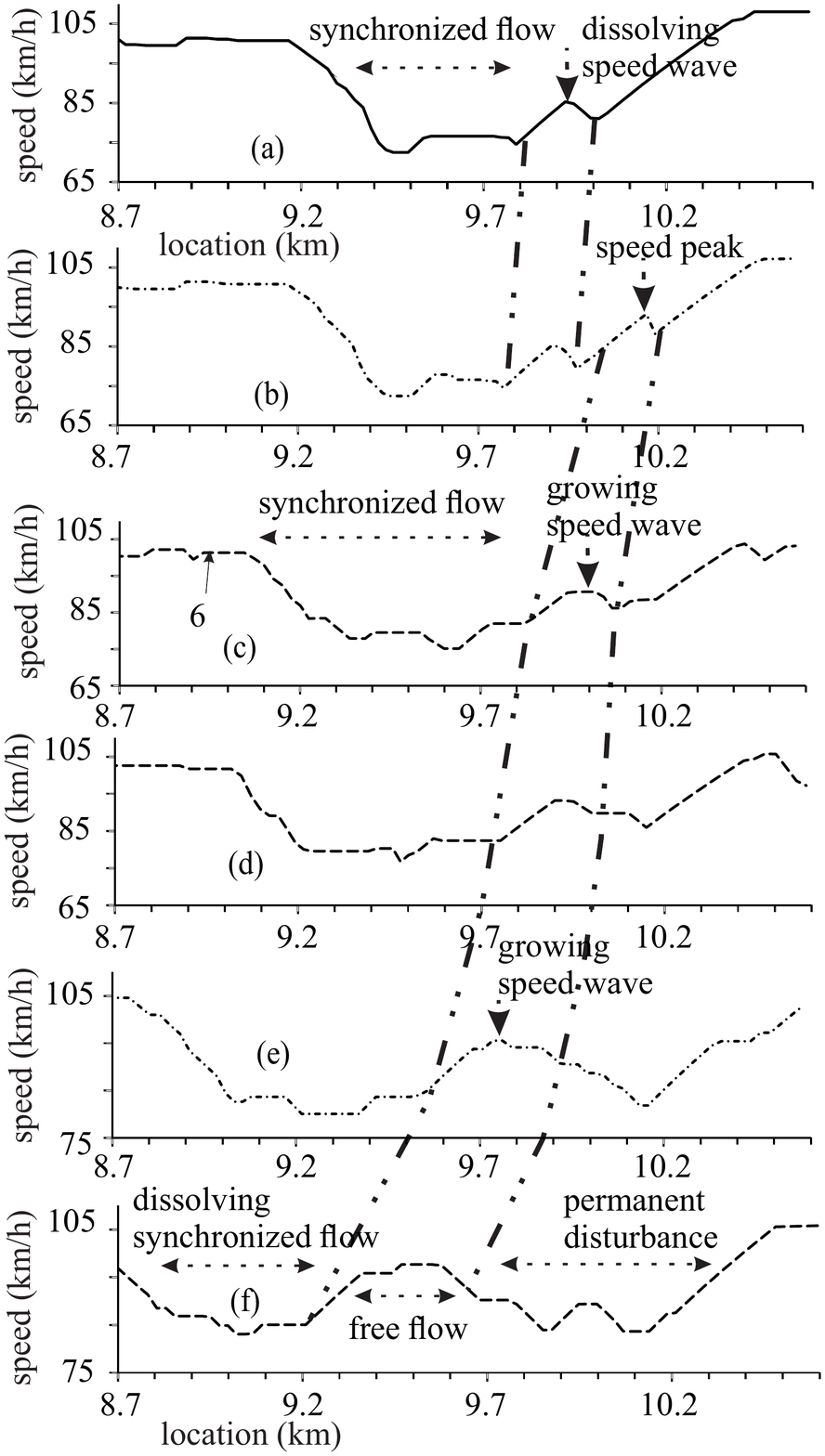}
 \end{center}
\caption{Microscopic vehicle speeds  as  
road location-functions related to Fig.~\ref{FSF_KKl_traj1_F_A} (e): Some of the    vehicles       
 moving at different times (that increase from (a) to (f), respectively, within a time interval between vehicles 5 and 8 shown in Fig.~\ref{FSF_KKl_traj1_F_A} (e)) propagate
  through the emergent synchronized flow  that is   marked
by dashed-dotted lines in Fig.~\ref{FSF_KKl_traj1_F_A} (d, e).
Dissolving and growing speed waves of increase in speed within the emergent synchronized flow
are marked by bold dashed-dotted curves
labeled by $\lq\lq$dissolving speed wave"
and
$\lq\lq$growing speed wave", respectively. 
Vehicle 6 in (c) is the same as
vehicle 6 marked in Fig.~\ref{FSF_KKl_traj1_F_A} (e).  
}
\label{FSF_traj6_KKl_F}
\end{figure}

As found in Secs.~\ref{Nuc_Sec} and~\ref{Gen_S} based on simulations with the KKSW CA model, simulations with the Kerner-Klenov model show also that
an S$\rightarrow$F instability tries to prevent an F$\rightarrow$S transition in free flow at the bottleneck as follows (Figs.~\ref{FSF_KKl_traj1_F_A}  
and~\ref{FSF_traj6_KKl_F}). 
  
(i) When the on-ramp inflow $q_{\rm on}$ is switched on ($t> 0$ in Fig.~\ref{FSF_KKl_traj1_F_A} (a, b)),  
vehicles that merge from the on-ramp onto the main road 
 cause a speed disturbance of decrease in speed in  free flow on the main road in a neighborhood of the bottleneck. The following vehicles have to
decelerate while adapting their speed a smaller speed within the disturbance. Due to this speed adaptation effect,
  synchronized flow emerges
 on the main road upstream at the bottleneck. See an example of the beginning of a such F$\rightarrow$S transition
  at   time instant    $t_{\rm FS}$ in Fig.~\ref{FSF_KKl_traj1_F_A} (d).
 The mean  speed  in this emergent synchronized flow is the smaller, the larger the initial speed disturbance of decrease in speed in  free flow.
 
 (ii) Within the downstream front of the emergent synchronized flow, speed peaks appear. Small speed peaks cause
 dissolving waves of increase in speed in the synchronized flow ($\lq\lq$dissolving
  speed wave" in Fig.~\ref{FSF_traj6_KKl_F} (a, b)). 
 When a large enough speed peak occurs, 
 the peak initiates a growing wave   of increase in speed within the synchronized flow
  ($\lq\lq$growing speed wave" in Fig.~\ref{FSF_traj6_KKl_F} (b--f)): At a time instant (labeled by $t_{\rm SF}$ in Fig.~\ref{FSF_KKl_traj1_F_A} (d))
 an S$\rightarrow$F instability is realized at the bottleneck. This S$\rightarrow$F instability
  destroys the emergent synchronized flow. As a result, the region of synchronized flow
 dissolves and free flow recovers at the bottleneck.
In accordance with Sec.~\ref{Mic_FSF_S}, the sequence of the emergence of the synchronized flow  
 (the beginning of an F$\rightarrow$S  transition)  with the subsequent S$\rightarrow$F instability
can be considered F$\rightarrow$S$\rightarrow$F transitions at the bottleneck
(Fig.~\ref{FSF_KKl_traj1_F_A} (c--f); compare with
 Fig.~\ref{FSF_KKSW_traj1_F_A} (b--e)). Due to many sequences of F$\rightarrow$S$\rightarrow$F transitions,
local permanent speed disturbance is realized in free flow at the bottleneck (time interval $0<t<T^{\rm (B)}$ in Fig.~\ref{FSF_KKl_traj1_F_A} (a, b);
compare with Fig.~\ref{FSF_KKSW_traj1_F_A} (a)). 

(iii) As long as   F$\rightarrow$S$\rightarrow$F transitions occur, no traffic breakdown (F$\rightarrow$S  transition)
with the subsequent formation of congested pattern is realized at the bottleneck (time interval $0<t<T^{\rm (B)}$ in Fig.~\ref{FSF_KKl_traj1_F_A} (a, b); compare with
 Fig.~\ref{FSF_KKSW_traj1_F_A} (a)) during time interval $0<t<T^{\rm (B)}_{1}$).

(iv) The S$\rightarrow$F instability exhibits the nucleation nature.
Therefore, there can be a random time instant $t=T^{\rm (B)}$ at which 
{\it no} S$\rightarrow$F instability occurs that can prevent the development of an F$\rightarrow$S  transition. 
In this case, the F$\rightarrow$S  transition leads to the formation of the congested pattern (WSP in Fig.~\ref{FSF_KKl_traj1_F_A} (a, b)
  at $t>T^{\rm (B)}$; compare with Fig.~\ref{FSF_KKSW_traj1_F_A} (a) at $t>T^{\rm (B)}_{1}$).
  
Thus as simulations with the KKSW CA model (Secs.~\ref{Inst_Mic_Over_I_S}--\ref{Gen_S}), simulations with the Kerner-Klenov model (Figs.~\ref{SFS_dis_onramp_I_KKl}--\ref{FSF_KKl_traj1_F_A}) prove that
small disturbances of decrease in speed  in free flow at the bottleneck are destroyed through the S$\rightarrow$F instability. In contrast, great enough
disturbances of decrease in speed in free flow cannot be destroyed resulting in an F$\rightarrow$S  transition with
 the formation of the congested pattern at the bottleneck. This explains why
through the nucleation character of the S$\rightarrow$F instability caused by the over-acceleration effect, 
free flow at the bottleneck is in a metastable state with respect to the F$\rightarrow$S  transition
and there is a random time delay $T^{\rm (B)}$ to this F$\rightarrow$S  transition.

 \subsection{Conclusions  \label{Con_S}}
 
The  
 S$\rightarrow$F instability exhibits the following general microscopic features, which are qualitatively identical ones
 in simulations with the KKSW CA   and Kerner-Klenov stochastic traffic flow models in the framework of the three-phase  theory.
 
 \subsubsection{Summary of nucleation features of S$\rightarrow$F instability  \label{Features_S}}

(i) An initial speed disturbance of increase in  speed within synchronized flow (S) at the bottleneck can transform into a growing speed wave of increase in   speed
(growing acceleration wave)
that propagates upstream within synchronized flow and leads to free flow (F) at the bottleneck.    This S$\rightarrow$F instability is caused by the over-acceleration effect.

(ii) The S$\rightarrow$F instability can occur, if there is a finite time delay in   over-acceleration.

(iii) Due to the S$\rightarrow$F instability, the downstream front of the initial synchronized flow begins to move upstream from the bottleneck, while free flow appears at the bottleneck.

(iv) In simulations, the initial speed disturbance of increase in speed that initiates the S$\rightarrow$F instability at the bottleneck
occurs at the downstream front of synchronized flow. We call the initial speed disturbance as $\lq\lq$speed peak".

(v) There can be many speed peaks with random amplitudes that occur randomly over time at the downstream front of synchronized flow. Only when a large enough speed
peak appears, the S$\rightarrow$F instability occurs. Speed peaks of smaller amplitude cause dissolving speed waves of increase in  speed (dissolving acceleration waves)
in synchronized flow: All these waves dissolve over time while propagating upstream within synchronized flow. As a result, the synchronized flow
persists at the bottleneck. Thus, the S$\rightarrow$F instability exhibits the nucleation nature.

  \subsubsection{S$\rightarrow$F instability as origin
  of nucleation nature of traffic breakdown \label{Con_S_2}}

The S$\rightarrow$F instability in synchronized flow at the bottleneck   governs   traffic breakdown (i.e., F$\rightarrow$S transition) resulting in
the formation of a congested pattern at the bottleneck as follows. 

(i) {\it A sequence of F$\rightarrow$S$\rightarrow$F transitions that interrupts the formation of a congested pattern at the bottleneck.}
When   an F$\rightarrow$S transition begins to develop, i.e., the upstream front of    synchronized flow  
    begins  to propagate upstream from the bottleneck,  an S$\rightarrow$F instability   can randomly occur.
Due to the S$\rightarrow$F instability,  free flow appears at the bottleneck. As a result,  
the downstream front of the   synchronized flow departs upstream  from the bottleneck.
 In its turn, this results in the dissolution of the synchronized flow, i.e., in the interruption of the formation of a congested pattern due to
 the F$\rightarrow$S transition. We call this effect as the sequence of F$\rightarrow$S$\rightarrow$F transitions.

(ii) {\it Metastability of free flow with respect to traffic breakdown (F$\rightarrow$S  transition) and a random time delay to traffic breakdown.}
 There can be many sequences of F$\rightarrow$S$\rightarrow$F transitions. Each of them  interrupts
  the formation of a congested pattern at the bottleneck. This explains the existence of a
time delay of traffic breakdown: Rather than the congested pattern appears at the bottleneck, the 
sequences of F$\rightarrow$S$\rightarrow$F transitions result in a narrow  region of  decrease in speed in free flow
 localized  at the bottleneck (called   as a $\lq\lq$permanent speed disturbance" in free flow at the bottleneck).
 The time delay of traffic breakdown (F$\rightarrow$S transition) $T^{\rm (B)}$  is a random value: 
 There can be a time instant $T^{\rm (B)}$ at which, after an F$\rightarrow$S transition begins to develop,
  there is
  no  S$\rightarrow$F instability that can prevent the subsequent development of the F$\rightarrow$S  transition.
 This F$\rightarrow$S  transition leads to the formation of a congested pattern at the bottleneck.
    
    Microscopic qualitative features of the S$\rightarrow$F instability    
     exhibit general character: These features are independent
 on   specific properties of a stochastic   traffic flow  model that incorporates hypotheses of
  the three-phase  theory.

An empirical evidence of  S$\rightarrow$F transitions at highway bottlenecks have been proven in~\cite{Kerner2002A}.
However, real field traffic data studied in~\cite{Kerner2002A} (as well as in all other publications known to the author) are {\it macroscopic} traffic data.
To prove the microscopic theory developed in this article with real field traffic data,
  measurements of {\it microscopic (single-vehicle)} spatiotemporal
data (e.g., vehicle trajectories) of almost all vehicles moving in free and synchronized flows in a neighborhood of a highway bottleneck are required. Unfortunately, such empirical microscopic traffic data
is not currently available. Therefore, a microscopic empirical study of traffic flow will be a very interesting task for further investigations of traffic flow.

\appendix
\section{Kerner-Klenov model for single-lane road with on-ramp bottleneck \label{App}}

In this Appendix, we present a discrete version of the Kerner-Klenov stochastic three-phase traffic flow model for single-lane road with on-ramp bottleneck~\cite{KKl2009A}
used in simulations shown in Figs.~\ref{SFS_dis_onramp_I_KKl}--\ref{FSF_traj6_KKl_F} (Sec.~\ref{KKSW_KKl_S}). In the model (Tables~\ref{table_CA}--\ref{table1}),
 index $n$ corresponds 
to the discrete time $t_{\rm n}=\tau n, \ n=0,1,...$, 
$v_{n}$ is the vehicle speed at time step $n$, $a$ is the maximum acceleration,
$\tilde v_{n}$ is the vehicle speed  without  speed fluctuations $\xi_{n}$, the lower index $\ell$  
marks variables related to the preceding vehicle, $v_{{\rm s}, n}$ is a safe speed at time step $n$,
$v_{\rm free}$ is the maximum speed in free flow,
  $\xi_{n}$ describes   speed fluctuations;
 $v_{{\rm c},n}$ is a desired speed;
all vehicles have the same length $d$ that includes
the mean space gap between vehicles within a wide moving jam where the speed is  zero.
In the model, discretized space coordinate with a small
enough value of the discretization cell $\delta x$ is used. Consequently,
the vehicle speed and acceleration discretization intervals are $\delta v=\delta x/ \tau$ and $\delta a=\delta v/ \tau$, respectively.
 In the model of an on-ramp bottleneck (Table~\ref{table1};
see explanations of model parameters in Fig.~16.2 (a)  of~\cite{KernerBook}),
superscripts    $+$   and  $-$  in variables, parameters, and functions 
denote the preceding vehicle and the trailing vehicle 
on the main road into which the vehicle moving in the on-ramp lane wants to merge.
 Initial and boundary conditions are the same as that explained in Sec.~16.3.9 of~\cite{KernerBook}. 
 Model parameters are presented in Tables~\ref{table_parameters} and~\ref{table_parameters_bottlenecks}.
 The physics of the model has been explained in~\cite{KKl2009A}.

\begin{table}
\caption{Discrete stochastic model~\cite{KKl2009A}}
\label{table_CA}
\begin{tabular}{|l|}
\hline
\multicolumn{1}{|c|}{
$v_{ n+1}=\max(0, \min({v_{\rm free}, \tilde v_{ n+1}+\xi_{ n}, v_{ n}+a
\tau, v_{{\rm s},n} })),$ 
}\\
\multicolumn{1}{|c|}{
$x_{n+1}= x_{n}+v_{n+1}\tau$,
}\\
\multicolumn{1}{|c|}{
$\tilde v_{n+1}=\max(0, \min(v_{\rm free},  v_{{\rm s},n}, v_{{\rm c},n})),
$
}\\
\multicolumn{1}{|c|}{
$v_{{\rm c},n}=\left\{\begin{array}{ll}
v_{ n}+\Delta_{ n} &  \textrm{at $g_{n} \leq G_{ n}$,} \\
v_{ n}+a_{ n}\tau &  \textrm{at $g_{n}> G_{ n}$}, \\
\end{array} \right.$
} \\
\multicolumn{1}{|c|}{
$\Delta_{ n}=\max(-b_{ n}\tau, \min(a_{ n}\tau, \ v_{ \ell,n}-v_{ n})),$
} \\
\multicolumn{1}{|c|}{
 $g_{n}=x_{\ell, n}-x_{n}-d$,
} \\
$v_{\rm free}$, $a$, $d$, and $\tau$ are constants. \\
\hline
\end{tabular}
\end{table}
\vspace{1cm} 

\begin{table}
\caption{Functions in discrete stochastic model I: Stochastic time delay of acceleration and
deceleration}
\label{table_CA1}
\begin{center}
\begin{tabular}{|l|}
\hline
\multicolumn{1}{|c|}{$a_{n}=a  \Theta (P_{\rm 0}-r_{\rm 1})$, \
$b_{n}=a  \Theta (P_{\rm 1}-r_{\rm 1})$,} \\
\multicolumn{1}{|c|}{
$P_{\rm 0}=\left\{
\begin{array}{ll}
p_{\rm 0} & \textrm{if $S_{ n} \neq 1$} \\
1 &  \textrm{if $S_{ n}= 1$},
\end{array} \right.
\quad
P_{\rm 1}=\left\{
\begin{array}{ll}
p_{\rm 1} & \textrm{if $S_{ n}\neq -1$} \\
p_{\rm 2} &  \textrm{if $S_{ n}= -1$},
\end{array} \right.$
}\\
\multicolumn{1}{|c|}{
$S_{ n+1}=\left\{
\begin{array}{ll}
-1 &  \textrm{if $\tilde v_{ n+1}< v_{ n}$} \\
1 &  \textrm{if $\tilde v_{ n+1}> v_{ n}$} \\
0 &  \textrm{if $\tilde v_{ n+1}= v_{ n}$},
\end{array} \right.$
}\\
$r_{1}={\rm rand}(0,1)$, $\Theta (z) =0$ at $z<0$ and $\Theta (z) =1$ at $z\geq 0$; \\
$p_{\rm 0}=p_{\rm 0}(v_{n})$, $p_{\rm 2}=p_{\rm 2}(v_{n})$  are speed functions,
 $p_{\rm 1}$ is constant. \\
\hline
\end{tabular}
\end{center}
\end{table}
\vspace{1cm} 
  
\begin{table}
\caption{Functions in discrete stochastic model II: Model speed fluctuations}
\label{table_CA2}
\begin{center}
\begin{tabular}{|l|}
\hline
\multicolumn{1}{|c|}{
$\xi_{ n}=\left\{
\begin{array}{ll}
\xi_{\rm a} &  \textrm{if  $S_{ n+1}=1$} \\
- \xi_{\rm b} &  \textrm{if $S_{ n+1}=-1$} \\
\xi^{(0)} &  \textrm{if  $S_{ n+1}=0$},
\end{array} \right.$
}\\
\multicolumn{1}{|c|}{$\xi_{\rm a}=a^{(\rm a)} \tau \Theta (p_{\rm a}-r)$, \
$\xi_{\rm b}=a^{(\rm b)} \tau \Theta (p_{\rm b}-r)$,} \\
\multicolumn{1}{|c|}{
$\xi^{(0)}=a^{(0)}\tau \left\{
\begin{array}{ll}
-1 &  \textrm{if $r\leq p^{(0)}$} \\
1 &  \textrm{if $p^{(0)}< r \leq 2p^{(0)}$ and $v_{n}>0$} \\
0 &  \textrm{otherwise},
\end{array} \right.$
}\\
$r={\rm rand}(0,1)$;
 $p_{\rm a}$, $p_{\rm b}$, $p^{(0)}$, 
 $a^{(0)}$, 
$a^{(\rm a)}$, $ a^{(\rm b)}$
are constants, \\
\hline
\end{tabular}
\end{center}
\end{table}
\vspace{1cm}

\begin{table}
\caption{Functions in discrete stochastic model III: Synchronization gap $G_{n}$ and safe speed $v_{{\rm s},n}$}
\label{table_CA3}
\begin{center}
\begin{tabular}{|l|}
\hline
\multicolumn{1}{|c|}{
$G_{n}=G(v_{n}, v_{\ell,n})$,
} \\
\multicolumn{1}{|c|}{
$G(u, w)=\max(0,  \lfloor k\tau u+  a^{-1}u(u-w) \rfloor),$
} \\
  $k>1$ is constant. \\
\multicolumn{1}{|c|}{
$v_{{\rm s},n}=
\min{(v^{\rm (safe)}_{ n},  g_{ n}/ \tau+ v^{\rm (a)}_{ \ell})},$
} \\
\multicolumn{1}{|c|}{
$v^{\rm (a)}_{\ell}=
\max(0, \min(v^{\rm (safe)}_{ \ell, n}, v_{ \ell,n}, g_{ \ell, n}/\tau)-a\tau),$
} \\
\multicolumn{1}{|c|}{
$v^{\rm (safe)}_{ n}=\lfloor v^{\rm (safe)} (g_{n}, \ v_{ \ell,n}) \rfloor,$ 
} \\
 $v^{\rm (safe)} (g_{n}, \ v_{ \ell,n}) $ is        taken  as that in~\cite{ach_Kra10}, 
\\ which is a solution of  the
 Gipps's equation~\cite{Gipps} \\
 \multicolumn{1}{|c|}{
$v^{\rm (safe)} \tau_{\rm safe} + X_{\rm d}(v^{\rm (safe)}) = g_{n}+X_{\rm d}(v_{\ell, n})$,
} \\
where   $\tau_{\rm safe}$
 is a safe time gap, \\
 \multicolumn{1}{|c|}{
$X_{\rm d} (u)=b \tau^{2} \bigg(\alpha \beta+\frac{\alpha(\alpha-1)}{2}\bigg)$,
} \\
\multicolumn{1}{|c|}{
$\alpha=\lfloor u/b\tau \rfloor$ and $\beta=u/b\tau-\alpha$ 
} \\
are the integer and  fractional parts  of $u/b\tau$, \\
respectively; 
$b$ is constant. \\
\hline
\end{tabular}
\end{center}
\end{table}
\vspace{1cm} 

\begin{table}
\caption{Models of vehicle merging at on-ramp bottleneck
that occurs when   a safety rule ($\ast$) {\it or} a safety rule  ($\ast \ast$) is satisfied 
}
\label{table1}
\begin{center}
\begin{tabular}{|l|}
\hline
\multicolumn{1}{|c|}{Safety rule ($\ast$):}\\
\multicolumn{1}{|c|}{
$\begin{array}{ll}
g^{+}_{n} >\min(\hat  v_{n}\tau , \ G(\hat  v_{n}, v^{+}_{n})), \\
g^{-}_{n} >\min(v^{-}_{n}\tau, \ G(v^{-}_{n},\hat  v_{n})),
\end{array} $
}\\
\multicolumn{1}{|c|}{
$\hat v_{n}=\min(v^{+}_{n},  \ v_{n}+\Delta v^{(1)}_{r}),$
} \\
 $\Delta v^{(1)}_{r}>0$ is constant.\\
\hline
\multicolumn{1}{|c|}{Safety rule ($\ast \ast$):}\\
\multicolumn{1}{|c|}{
$x^{+}_{n}-x^{-}_{n}-d > \lfloor  \lambda_{\rm b} v^{+}_{n} +d \rfloor,$
}\\
\multicolumn{1}{|c|}{
$\begin{array}{ll}
x_{n-1}< x^{\rm (m)}_{n-1} \  \textrm{and} \
 x_{n} \geq x^{\rm (m)}_{n} \\
\ \textrm{or} \
x_{n-1} \geq x^{\rm (m)}_{n-1} \  \textrm{and} \
 x_{n} < x^{\rm (m)}_{n},
\end{array}$
}\\
\multicolumn{1}{|c|}{
$x^{\rm (m)}_{n}=\lfloor (x^{+}_{n}+x^{-}_{n})/2 \rfloor,$
}\\
$\lambda_{\rm b}$ is constant. \\
\hline
\multicolumn{1}{|c|}{Parameters after vehicle merging:}\\
\multicolumn{1}{|c|}{$v_{n}=\hat v_{n},$}\\
\multicolumn{1}{|c|}{under the rule ($\ast $): $x_{n}$  maintains the
same,}\\
\multicolumn{1}{|c|}{under the rule ($\ast \ast$): $x_{n} = x^{\rm
(m)}_{n}$.}\\
\hline
\multicolumn{1}{|c|}{Speed adaptation before vehicle merging}\\
\multicolumn{1}{|c|}{
$v_{{\rm c},n}=\left\{\begin{array}{ll}
v_{ n}+\Delta^{+}_{ n} &  \textrm{at $g^{+}_{n} \leq G(v_{n}, \hat
v^{+}_{n})$,} \\
v_{ n}+a_{ n}\tau &  \textrm{at $g^{+}_{n}>G( v_{n}, \hat
v^{+}_{n})$}, \\
\end{array}\right. $
}\\
\multicolumn{1}{|c|}{
$\Delta^{+}_{ n}=\max(-b_{ n}\tau, \min(a_{ n}\tau, \ \hat v^{+}_{n}-v_{
n})),$
}\\
\multicolumn{1}{|c|}{
$\hat v^{+}_{n}=\max(0, \min(v_{\rm free}, \  v^{+}_{n}+\Delta
v^{(2)}_{r})),$
}\\
$\Delta v^{(2)}_{r}$ is  constant. \\
\hline
\end{tabular}
\end{center}
\end{table}
\vspace{1cm}

\begin{table}
\caption{Model parameters   used in Figs.~\ref{SFS_dis_onramp_I_KKl}--\ref{FSF_traj6_KKl_F}: Vehicle motion in road lane}
\label{table_parameters}
\begin{center}
\begin{tabular}{|l|}
\hline
$\tau_{\rm safe}   = \tau=$ 1, $d = 7.5 \  \rm m/\delta x$, \\
$\delta x=$ 0.01 m, $\delta v= 0.01 \  {\rm ms^{-1}}$, $\delta a= 0.01 \  {\rm ms^{-2}}$, \\
$v_{\rm free}= 30 \ {\rm ms^{-1}}/\delta v$, $b = 1 \ {\rm ms^{-2}}/\delta a$, $a=$ 0.5 ${\rm ms^{-2}}/\delta a$, \\
$k=$ 3, $p_{1}=$ 0.3,  $p_{b}=   0.1$,  $p_{a}=   0.17$, $p^{(0)}= 0.005$, \\
$p_{\rm 2}(v_{n})=0.48+ 0.32\Theta{( v_{n}-v_{21})}$, \\
$v_{01} = 10 \ {\rm ms^{-1}}/\delta v$, $v_{21} = 15 \ {\rm ms^{-1}}/\delta v$, \\
$p_{\rm 0}(v_{n})=0.575+ p_{\rm 01}\min{(1, v_{n}/v_{01})}$, \\
  $a^{(0)}= 0.2a$,   $a^{(\rm a)}= a$,   
$a^{(\rm b)}= a$; \\   
   $p_{\rm 01}=$ 0.205 
in Fig.~\ref{SFS_dis_onramp_I_KKl} and 0.125 in Figs.~\ref{FSF_KKl_traj1_F_A} and~\ref{FSF_traj6_KKl_F}. \\   
   \hline
\end{tabular}
\end{center}
\end{table}
\vspace{1cm}

\begin{table}
\caption{Parameters of model of on-ramp bottleneck used in Figs.~\ref{SFS_dis_onramp_I_KKl}--\ref{FSF_traj6_KKl_F}}
\label{table_parameters_bottlenecks}
\begin{center}
\begin{tabular}{|l|}
\hline
$\lambda_{\rm b}=$ 0.75, 
   $v_{\rm free \ on}=22.2 \ {\rm ms^{-1}}/\delta v$,  \\
   $\Delta v^{\rm (2)}_{\rm r}=$ 5 \  ${\rm ms^{-1}}/\delta v$, 
   $L_{\rm r}=1 \ {\rm km}/\delta x$,  \\ $\Delta v^{\rm (1)}_{\rm r}=10 \ {\rm ms^{-1}}/\delta v$,  
   $L_{\rm m}=$ 0.3 \    ${\rm km}/\delta x$. \\
   \hline
\end{tabular}
\end{center}
\end{table}
\vspace{1cm}

 {\bf Acknowledgments:}
 We thank our partners for their support in the project $\lq\lq$UR:BAN - Urban Space: User oriented assistance systems and network management", funded by the German Federal Ministry of Economics and Technology. 
I thank Sergey Klenov for discussions and help in simulations.

\end{document}